\documentclass[applsci,review,submit,moreauthors,pdftex]{Definitions/mdpi} 

\def\nue{\ensuremath{\nu_{e}\ }}
\def\nubare{\ensuremath{\overline{\nu}_{e}}\ }

\def\numu{\ensuremath{\nu_{\mu}\ }}
\def\nubarmu{\ensuremath{\overline{\nu}_{\mu}}}
\def\nubartau{\ensuremath{\overline{\nu}_{\tau}}}

\def\nutau{\ensuremath{\nu_{\tau}\ }}

\newcommand{\numunue}{\ensuremath{\nu_\mu \rightarrow \nu_e\,}}

\newcommand{\nubarmunubare}{\ensuremath{\overline{\nu}_\mu \rightarrow \overline{\nu}_e\,}}
\newcommand{\nova}{NO$\nu$A$\,$}

\usepackage{textcomp}
\usepackage{amsmath,amssymb}
\firstpage{1} 
\pubvolume{1}
\issuenum{1}
\articlenumber{0}
\pubyear{2020}
\copyrightyear{2020}
\history{Received: date; Accepted: date; Published: date}

\pdfoutput=1

\Title{Design and diagnostics of high-precision accelerator neutrino beams}

\Author{N. Charitonidis$^{1}$\orcidN{}, A. Longhin$^{2}$\orcidB{}, M.~Pari$^{1,2}$\orcidC{} E.G.  Parozzi$^{1,3}$\orcidE{} and F. Terranova$^{3}$*\orcidD{}}

\AuthorNames{Nikolaos Charitonidis, Andrea Longhin, Elisabetta Parozzi, Francesco Terranova}

\address{%
$^{1}$ \quad CERN, 1211 Geneva 23, Switzerland \\
$^{2}$ \quad Dep. of Physics, University of Padova and INFN, Padova, Italy;\\
$^{3}$ \quad Dep. of Physics, University of Milano-Bicocca and INFN, Milano, Italy; 
}

\corres{Corresponding Author: F. Terranova. E-mail: francesco.terranova@cern.ch}

\abstract{Neutrino oscillation physics has entered a new precision era, which poses major challenges to the level of control and diagnostics of the neutrino beams. In this paper, we review the design of high-precision beams, their current limitations, and the latest techniques envisaged to overcome such limits. We put emphasis on ``monitored neutrino beams'' and advanced diagnostics to determine the flux and flavor of the neutrinos produced at the source at the per-cent level. We also discuss ab-initio measurements of the neutrino energy -- i.e.~measurements performed without relying on the event reconstruction at the $\nu$ detector -- to remove any flux induced bias in the determination of the cross sections.}


\begin{document}

\section{Introduction}
\label{sec:introduction}

Accelerator neutrino beams played a pivotal role in unveiling the electroweak sector of the Standard Model and in the discovery of Neutrino Oscillations~\cite{Giganti:2017fhf}. Unlike natural neutrino sources, they offer a superior degree of control on the momentum, direction and flavor of the neutrinos at the source and therefore, are the ideal facilities for the precision era of neutrino oscillation physics~\cite{Mezzetto:2020jka}. Before the discovery of neutrino oscillations, hadron beam diagnostics was a useful tool to determine the beamline performance, the integrated intensity and the flux variation over time. After the discovery of non-zero $\theta_{13}$ in 2012~\cite{Tanabashi:2018oca}, it is evident that these facilities are the optimal tools to investigate lepton mixing with unprecedented precision. Up to now, accelerator neutrino experiments provided a wealth of information on neutrino mixing and masses but they face unprecedented challenges in terms of precision: the next generation of experiments for the study of the missing parameters (the CP violating phase, the mass hierarchy and the octant of $\theta_{23}$~\cite{Esteban:2020cvm}) employs multi-MW proton beams and very massive detectors ($>40,000$~tonnes) located in underground laboratories, hundreds of km far from the beam source. These experiments are called {\em long-baseline} experiments because the beam-to-detector distance is very large ($\gg 10$~km). For the first time in the history of neutrino physics, these facilities will be no more limited by the event statistics -- thanks to the detector mass and beam power -- but by a large set of systematic uncertainties that must be controlled at the per-cent level to achieve our new physics goals~\cite{Huber:2007em,Coloma:2012ji}.   
Beam diagnostics is then prominent for the success of the next generation of experiments and, in particular, for DUNE~\cite{Abi:2020wmh} and Hyper-Kamiokande~\cite{Abe:2018uyc} (HK) that will be in data taking in less than a decade. 

Similarly, novel designs and advanced diagnostics are important tools to reduce the systematic uncertainties of neutrino cross-sections, which are still at $10$-$30$\% level in the few-GeV neutrino energy range~\cite{katori2018}. 
The cross sections are measured by dedicated experiments where, unlike DUNE or HK, the neutrino detector is located at a short distance from the source to ensure that no oscillation takes place
({\em short-baseline} experiments). The neutrino cross sections are extracted by the number of interacting neutrinos in the detector, which is given by:
\begin{equation}
    N(E_R) \sim \int dE \ \sigma_f(E) \epsilon_f(E\rightarrow E_R) \phi_f(E)
    \label{eq:cross_section}
\end{equation}
where $N$ is the number of neutrino interactions observed in the detector for a reconstructed neutrino energy $E_R$, $\sigma_f(E)$ is the neutrino cross section at the true energy $E$ for a given flavor ($f= \nue, \numu, \nutau, \nubare, \nubarmu, \nubartau$), $\epsilon_f(E\rightarrow E_R)$ is the probability to detect a neutrino with energy $E$ and reconstruct it at the energy $E_R$, and $\phi_f(E)$ is the neutrino flux of that flavor expressed in neutrinos/m$^2$s. The coefficient that links the left and right side of Eq.~\ref{eq:cross_section} is the number of scattering centers crossed by the neutrinos and is, therefore, very well known from the detector mass.

Despite the complexity of the reconstruction of neutrino interactions, the leading source of systematics on the cross-sections is the flux $\phi_f(E)$, which is generally known with a precision worse than $10$\%. Beam diagnostics is nearly the only tool we have at our disposal to reduce substantially such uncertainty.

In general, precision neutrino physics is driving a revision of beam designs and boosting the field of high-precision beam diagnostics~\cite{NBI}, a branch of applied physics that has been perfected in colliders but overlooked in neutrino beams. This review focuses on the design and control of accelerator neutrino beams looking at the three most important applications of these facilities:

\begin{itemize}
\item Neutrino beams with unprecedented precision on $\phi_f(E)$ for the measurement of the \nue  and \nubare cross sections. These data are essential for DUNE and HK because long-baseline experiments study the oscillation of $\nu_\mu$ into $\nu_e$ and its CP-conjugate $\bar{\nu}_\mu \rightarrow \bar{\nu}_e$. They thus need an exquisite precision in $\sigma_{\nu_e}(E)$ and $\sigma_{\bar{\nu}_e}(E)$.
\item Cross sections for the study of the {\em disappearance channel}\footnote{A disappearance channel is a $\nu_f \rightarrow \nu_f$ oscillation, where the number of observed $\nu_f$ is smaller that the $\nu_f$ at the source because some of them have changed flavor. An appearance channel is a $\nu_f \rightarrow \nu_{f'}$ oscillation where a new flavor $f'$ appears into the detector.} $\nu_\mu \rightarrow \nu_\mu$ and $\bar{\nu}_\mu \rightarrow \bar{\nu}_\mu$ in long-baseline experiments. This measurement is essential to determine the mixing parameters and must be combined with the above-mentioned $\nu_e$ {\em appearance channels} (\numunue and \nubarmunubare) to disentangle effects due to CP violation from effects due to the neutrino mass hierarchy and (if any) deviations from the Standard Model~\cite{Escrihuela:2016ube}. In this case, a 1\% level measurement of $\sigma_{\nu_\mu}(E)$ and $\sigma_{\bar{\nu}_{\mu}}(E)$ is mandatory.
\item The knowledge of the absolute flux is less critical for long-baseline oscillation experiments than for short-baseline cross section experiments because DUNE and HK use ancillary detectors (``Near Detector'', ND) located near the source to estimate the incoming rate of neutrinos. At the same time, the absolute fluxes $\phi_{\numu}$ and $\phi_{\nubarmu}$ must be known with $<5$\% precision because the Near Detector samples a flux that is not exactly the same as the flux reaching the ``Far Detector'' (FD), which is located far from the source and designed to observe the oscillations. The imperfections in the flux cancellation arising by the ND-versus-FD comparison cannot be neglected any more in modern facilities.
Similar considerations hold for the size of the wrong-flavor contamination (e.g.~\nue at the source) that can pollute the $\numunue$ measurement at the FD.
\end{itemize}

In this work, we will show that high-precision cross section measurements need short-baseline  experiments with a dedicated design for the beamline and tailor-made diagnostics. For instance, in the {\em monitored neutrino beams}~\cite{Longhin:2014yta},  the flux is derived most straightforwardly by measuring the charged leptons produced in the decay volume of the beam. These leptons are directly proportional to the number of neutrinos created at the source. 

On the other hand, the design of the Near Detector in long-baseline beams must be tailored to measure the flux and wrong-flavor contamination at a few per-cent level. Spectral information on $\phi_f(E)$ is very useful to mitigate the systematics arising from an imperfect ND-versus-FD cancellation.  This task is substantially eased if the corresponding beamlines are complemented by diagnostic tools inherited from monitored neutrino beams and if the cross sections are known with a precision of $\sim$1\%.  

\section{Accelerator neutrino beams}
\label{sec:anb}

We introduce here the main elements of an accelerator neutrino beam. Each of these elements is being re-considered and redesigned when precision beams are considered, due to their special requirements on the flux measurements.
The first systematic studies on accelerator neutrino beams were presented in 1965~\cite{CERNconf65}, in a seminal CERN conference devoted to neutrinos. In ``conventional'' neutrino beams, the neutrinos are created by the decay of pions or kaons produced by the interaction of a high-energy proton beam on a fixed target. 
Conventional beams are mainly sources of $\numu$ and $\nubarmu$, which are produced by $\pi^\pm \rightarrow \mu^\pm \numu (\nubarmu)$. Electron-neutrinos are always subdominant because they are created either by kaons $K^\pm \rightarrow e^\pm \pi^0 \nue (\nubare)$ or by the decay in flight of $\mu^\pm$, like $\pi^+ \rightarrow \mu^+ \numu \rightarrow ( e^+ \nue \nubarmu) \ \numu$. The readers can find  an extensive review of these beams  in Ref.~\cite{kopp2006}.

\subsection{Narrow and wide-band conventional beams}
\label{sec:narrow_wide}

\begin{figure}[t]
\centering
\includegraphics[scale=0.4]{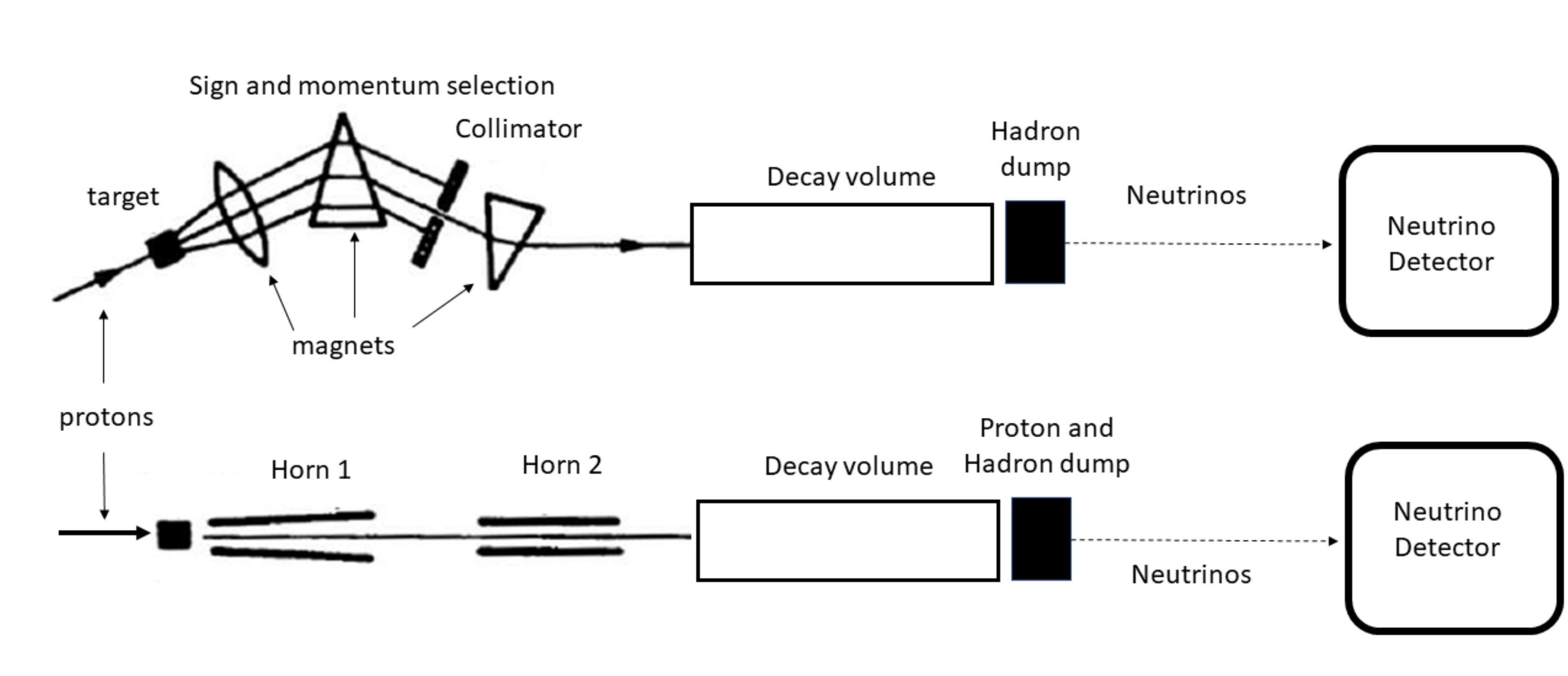}
    \caption{(top) A typical layout for a narrow-band beam. After the target, a set of quadrupoles, dipoles and collimators (slits) selects the sign and momentum and steers the secondaries toward the decay volume. All charged particles but neutrinos are stopped at the hadron dump located before the neutrino detector. The {\em baseline} of the facility is the target to neutrino detector distance. (bottom) A wide-band beam where the focusing and momentum selection is performed by a set of horns (in the figure, ``Horn 1'' and ``Horn 2''). The second horn is often called ``reflector''. Note that a horn can be inserted also in a narrow-band beam, generally between the target and the first quadrupole.}
    \label{fig:narrow_wide}
\end{figure}

 The vast majority of neutrino beams designed in the past~\cite{Dore:2018ldz}, together with those that are currently in operation, are optimized to ensure the largest neutrino {\em flux } in a broad energy range. These beams are referred-to as {\em wide-band beams}, where the term ``band'' refers to the width of the momentum spectrum of the hadrons and therefore, produced neutrinos (see Fig.~\ref{fig:narrow_wide}). These beams have been the workhorse of major discoveries in oscillation physics because they compensate the tiny cross sections of the neutrinos at the GeV scale ($\sigma \simeq 10^{-38}$~cm$^2$) with a significant flux enhancement. 

The maximum bandwidth of a neutrino beam can be calculated from a {\em bare target experiment}, where all produced secondary particles and the tertiary neutrinos freely stream towards the decay volume or the detector. As an illustration, for the case of the pions, the Feynman parton model~\cite{Feynman:1969wa,Tanabashi:2018oca} predicts that the average angle of emission of the secondaries is given by:
\begin{equation}
    \theta \simeq \frac{\langle p_T \rangle }{p_\pi} \simeq \frac{280 \ \mathrm{MeV}}{p_\pi} = \frac{2}{\gamma} 
\end{equation}
where $p_T$ and $p_\pi$ is the transverse and total momentum of the pion.
This is because the parton momentum inside a nucleus is different from zero and the transverse momentum is Lorentz invariant. 
Therefore, $p_T$ traces the momentum distribution of the parton inside the nucleus independently of the Lorentz boost along the proton direction. Experimental data~\cite{Tanabashi:2018oca} suggest
$\langle p_T \rangle \simeq 280$~MeV and this quantity is scale-independent, i.e.~it does not depend on the longitudinal momentum of the secondary particle produced by the parton. The longitudinal particle momentum is described by:
\begin{equation}
    \frac{d^2 N}{d x_F d p_T} \simeq f(x_F) g(p_T)
\label{eq:parton_factorization}
\end{equation}
where $f$ and $g$ are empirical functions depending only on the  Feynman's $x$ ($x_F \simeq p_z/p$) and the transverse momentum $p_T$, respectively. $p$ is the momentum of the secondary particle (e.g.~the pion) and $p_z$ its longitudinal component. Equation~\ref{eq:parton_factorization} shows that the longitudinal effects are factorized with respect to the transverse momentum. $g(p_T)$ then dominates the spread of the secondary beam after the target.

A wide-band beam (Fig.~\ref{fig:narrow_wide}, bottom) employs the focusing given by the horns, described in Sec.~\ref{subsec:horn}, to produce a large acceptance beam. The subsequent decays of the pions and the kaons produce neutrinos over a broad energy range. The pion angle decreases with $\gamma$ and, in turn, the corresponding $\nu_\mu$ is collinear with the pion within an angle $\simeq 1/\gamma$. As a consequence, if the focusing selects high-energy particles, the neutrinos are well-collimated by the Lorentz boost but transverse focusing is very important for low energy wide-band beam.%

\begin{figure}[h]
\centering
\includegraphics[scale=0.45]{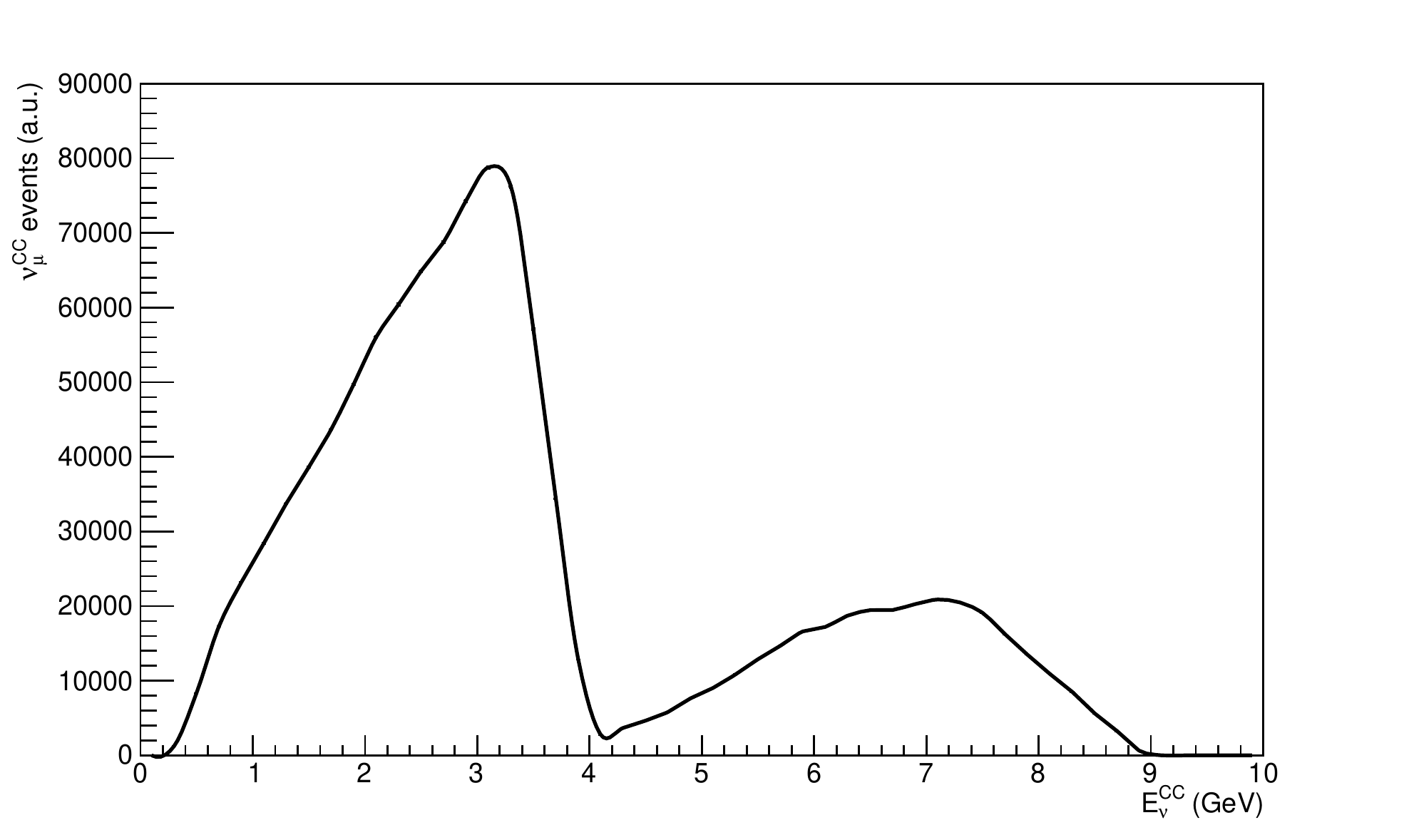}
    \caption{The spectrum expected for the NP06/ENUBET monitored neutrino beam. The high-energy peak is due to $\nu_\mu$ originating from kaons in the decay volume.}
    \label{fig:dichromatic}
\end{figure}

A {\em narrow-band beam } (Fig.~\ref{fig:narrow_wide}, top) offers a higher degree of precision at the expenses of a strong reduction of the flux. In this case, the momentum and charge of the hadrons (and, in turn, the produced neutrinos) are selected employing magnetic dipoles and slits. The typical narrow-band beam spectrum is called ``dichromatic''~\cite{kopp2006} because two distinct peaks are visible: the peak of the neutrinos coming from the selected pions and the peak of the neutrinos from the selected kaons (see Fig.~\ref{fig:dichromatic}). If the accepted momentum range -- the {\em momentum  bite} -- is very small, the width of the peak is dominated by the two-body decay of the pions or the kaons. It is virtually impossible to have a narrow $\nu_e$ peak in a dichromatic beam because the $\nu_e$'s arise from the three-body decay of the kaons ($K^+ \rightarrow e^+ \pi^0 \nue$) and may acquire a large transverse momentum if the neutrino is emitted in the direction opposite to the positron-$\pi^0$ pair.  In general, a narrow-band beam is an optimal choice for cross section measurements because it shrinks the intrinsic uncertainty on the neutrino energy and, therefore, on $\sigma_{\nu_\mu}(E)$. We will discuss in Sec.~\ref{sec:nboa} a technique to further reduce the energy spread without any flux loss.

On the other hand, high-intensity beams are in most cases wide-band beams and a determination of the neutrino energy must rely either on the energy reconstruction from the final state particles after a neutrino interaction or by the off-axis location of the detector, as explained in Sec.~\ref{sec:nboa}. In particular, DUNE is a wide-band beam with a mean energy of $\sim 3$~GeV while Hyper-Kamiokande employs a lower energy beam, whose energy spread is reduced by the off-axis location of the HK detector.
The first choice enhances the sensitivity to matter effects and, hence, to the neutrino mass hierarchy, while the latter is well suited to search for CP-violation.

\subsection{Non-conventional neutrino beams} 

Several ``non-conventional'' neutrino beam concepts have been developed over the years. In these cases, the neutrinos are produced by leptons or by the $\beta$-decays of partially ionized isotopes. The proposed {\em Neutrino Factories}~\cite{Choubey:2011zzq} would create, accelerate and store in a ring charged muons, exploiting the $\mu^{\pm} \rightarrow e^{\pm}+\nu_{e}({\bar{\nu}_{e})}+\bar{\nu}_{\mu}({\nu_{\mu})}$ decay channel. Here, the beam composition  is precisely known to be $50$\% $\nu_{e}(\bar{\nu_{e}})$ and $50$\% $\bar{\nu}_{\mu} (\nu_{\mu})$. If a Neutrino Factory were built, it would offer major advantages compared with  conventional beams.
Firstly, the number and polarization of the muons can be easily measured in the storage ring, offering superior control of flux and beam composition. Furthermore, the beam would be free of contamination because the neutrino flavors are uniquely determined by the muon decay products. Finally, a Neutrino Factory could be a powerful source of \nue and \nubare, unlike conventional beams where \nue and \nubare are always sub-dominant.  A systematic review of the physics of Neutrino Factories is available in~\cite{Geer1998,DeRujula:1998umv}. However, the technical challenges of a Neutrino Factory are tremendous. Due to the finite lifetime, the muons must be accelerated very quickly and the transverse momentum $p_T$ must be highly reduced ({\em muon cooling}) before injection into the storage ring~\cite{Parsa2000,Choubey:2011zzq,Bogomilov:2019kfj}.

Neutrinos can also be produced by beta decays of accelerated isotopes and generate a single-flavor beam ($\bar{\nu}_e$ or \nue for $\beta^-$ or $\beta^+$ decays, respectively) with excellent control of the flux. The neutrino energy spectrum  depends on the accelerator that boosts the isotope and the three-body kinematics of the $\beta$ decays. This facility, known as {\em Beta Beam}, has been proposed in 2002~\cite{Zucchelli:2002sa} but was never implemented despite a significant R\&D. Again, the technical challenges of creating, selecting and accelerating beta-unstable isotopes with high-rigidity have not been overcome, yet~\cite{book_beta_beams}.

\subsection{Neutrino production and hadronic cross-sections} 

As mentioned above, conventional beams are mainly  a source of $\nu_\mu$ with a relatively small contamination of $\nu_e$ from kaon decays or decay-in-flight of muons.
The relative weight of \nue from kaons and muons depends on the hadron energy, the length of the secondary hadron beamline and the decay volume . The smaller the energy, the higher the contribution of DIF because the kaon yield is suppressed. A long decay volume also enhances the DIF contribution due to the large difference between the muon and kaon lifetime. 

If the energy of the incoming proton beam is large, it is possible to have a small amount of $\nu_{\tau}$ coming from the tauonic decays of charmed mesons ($D_{s}$). In long-baseline experiments, the \nutau contamination is always negligible ($10^{-4}$) and can be enhanced only in beam dump experiments, where the neutrinos are detected at a very short distance from the beam dump~\cite{Kodama:2000mp}. In conclusion, conventional neutrino beams are powerful sources of \numu and $\bar{\nu}_\mu$ but the production of other flavors is strongly suppressed. They are particularly effective to study \numunue and \nubarmunubare oscillations but must be complemented by natural sources or non-conventional facilities if flavors different from \numu are needed.

Neutrino experiments require quite precise information on the production of the $\pi^{+}$,$\pi^{-}$,$K^{+}$, $K^{-}$ and $K^0_{L}$ mesons. This is mandatory since the originating hadron's momentum is directly correlated with the resulting neutrino energy by relativistic kinematics~\cite{CERNconf63}. 
Uncertainties in the hadron yields lead to the largest uncertainties in the neutrino fluxes. 
This crucial item will be discussed in Sec.~\ref{sec:hadron_yields}.

\subsection{Decay volume}

Every neutrino beam includes a {\em decay volume}
to allow for the charged mesons to decay  (see Fig.~\ref{fig:narrow_wide}). The decay volume is usually a cylindrical hollow tunnel, whose radius and length determine the number of neutrinos that reach the detector. The beam of secondary hadrons must have  a relatively small  divergence so that the particles do not impinge on the cylinder's wall before decaying. The cylinder ends up with a {\em hadron dump} that stops all charged hadrons from reaching the experiment.

The decay volume is generally not instrumented except for monitored neutrino beams, where the number of leptons produced by the kaon decays is recorded. Muons from $\pi^+ \rightarrow \mu^+ \nu_\mu$ impinge on the hadron dump and may be detected by positioning radiation-hard devices just after the dump (see Sec.~\ref{sec:hadron_dump}). 

\section{Extraction and monitoring of primary protons}
\label{sec:proton}

Since their early developments, neutrino beams have mostly relied on synchrotrons as the source of primary protons. For instance, the J-PARC Main Ring has been employed for T2K~\cite{t2k:overview}, the CERN Super Proton Synchrotron (SPS) for the West Area Neutrino Facility~\cite{Astier:2003gs} and CNGS~\cite{CNGSandOPERA}, the Fermilab (FNAL) Main Injector and Booster for the NuMI beamline~\cite{numi:beam} and MiniBooNE~\cite{miniboone:flux}, respectively.
Differently from linear accelerators, synchrotrons can reach high energies ($\mathcal{O}(10$-$100)$~GeV), 
and offer different proton extraction schemes that suit the maximum rates sustainable by the detectors and the diagnostics, and the features of the neutrino beamline.

The proton extraction methods employed for neutrino beamlines are based on two main schemes: fast and slow extraction.
\begin{itemize}
\item In the {\em fast extraction}, a kicker magnet is used to extract all protons stored in the machine in a time shorter than
a machine revolution period. Such time is typically in the $\mathcal{O}(1$-$10\text{~\textmu s})$ range. According to the experimental requirements, the protons can be shared into several macro-pulses. These pulses always have a fine structure that corresponds to the proton bunches produced
by the Radio Frequency (RF) cavities of the machine.

\item The {\em slow extraction} is driven by an unstable resonant motion of the particles in the transverse phase space, which is used to continuously spill the beam out of the machine over a long time (up to several seconds). This instability is controlled by acting on the machine optics (e.g.~``tune'') and non-linear elements (e.g.~sextupoles and/or octupoles).
\end{itemize}

The fast extraction is the method used by all neutrino beams currently in data taking: this scheme is driven by the use of
magnetic horns (see Sec.~\ref{subsec:horn}) as focusing devices for the secondaries, which result in high-intensity wide band beams. The flux of the secondary particles and, hence, of neutrinos, is enhanced by the large angular and momentum acceptance of the horns. Neutrino beams at the GeV scale, equipped by horns, benefit by an increase of acceptance of about one order of magnitude than a bare target experiment or a static beamline, i.e.~a beamline  where the focusing is performed only with static magnetic elements (dipoles and quadrupoles).
Among the drawbacks of a fast extraction, the first one is target fatigue, which is worsened by the fast and large instantaneous rate of the energy deposition (Sec.~\ref{sec:target}). 
The fast extraction also impacts the diagnostics and the instrumentation along the beamline, which has to cope with particle rates of tens of GHz per detector channel. Finally, it challenges the Near Detectors of long-baseline experiments, like DUNE, due to the pile-up of neutrino interactions. A fast extraction scheme is not viable in a monitored neutrino beam, where the leptons in the beamline have to be identified at a single-particle level. 

The low particle rate resulting from the long spills of the slow extraction can be compatible with an event-by-event detection process and extend the lifetime of the target (Sec.~\ref{sec:target}).
The main limitation of the slow extraction is the difficulty of using a magnetic horn, as the Joule heating from the long current pulses in the horn conductor can compromise the device beyond repair.
A remarkable example is the ESS$\nu$SB project~\cite{essnusb_14,essnusb_accumulator_1}, discussed in Sec.~\ref{subsec:horn}.

\begin{figure}[htb!]
    \centering
    \includegraphics[width=0.9\textwidth]{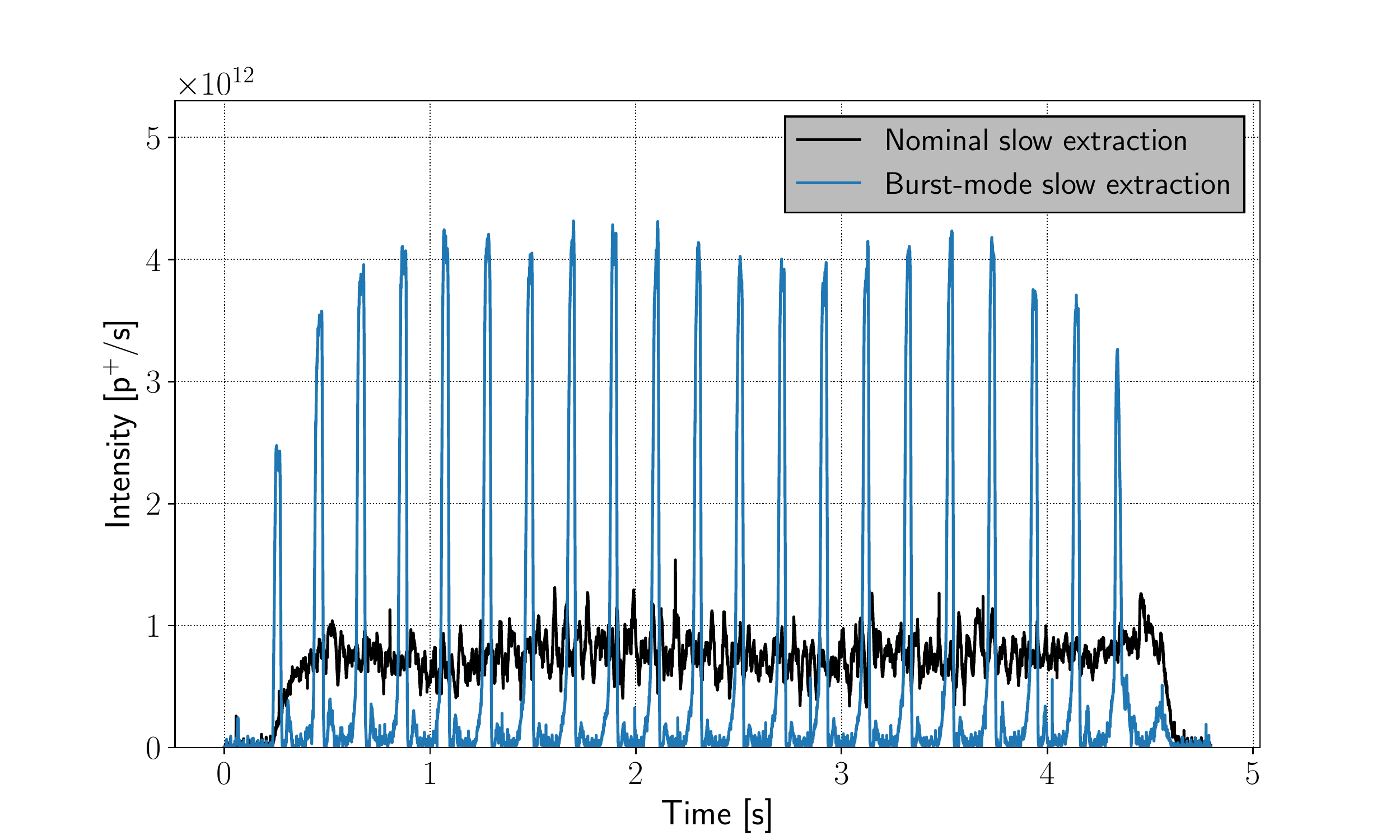}
    \caption{Comparison between a nominal slow extracted spill and a ``burst-mode slow extraction'' one. The spill profiles have been measured with a secondary emission monitor at the SPS, during dedicated machine tests. The same intensity is extracted in the two cases.}
    \label{fig:burstspill}
\end{figure}

Examples of fast and slow extraction implementations used for different types of neutrino beams, as well as the latest developments, can be drawn for instance from the CERN-SPS -- a machine that gave substantial contributions to neutrino physics since its early days.
The SPS can extract an intensity up to about $4.5\times10^{13}$~protons, with a particle momentum up to $450$~GeV/c. Recently, it was used to drive the CNGS neutrino
beam, with a fast extraction cycle providing two proton pulses $10.5$~\textmu s long of $\sim 2.4\times10^{13}$ protons each, separated by a $50$~ms time interval
and with a $5$~ns RF bunching structure~\cite{brennan:cngs,verena:cngs,cngs:firstbeam}. Similar pulses with a length of a few microseconds are obtained also for the neutrino beams at J-PARC~\cite{mr:overview,t2k:overview} and Fermilab~\cite{numi:beam,miniboone:flux}.
The nominal slow extraction of the SPS~\cite{verena:overview,verena:cose} is based on a third-integer resonance, and produces typically a few second long spill, which is delivered on multiple targets to produce the secondary particle beams serving the North Area (NA) fixed-target experiments. 
In this context, two narrow-band secondary hadron  beamlines have been recently built and commissioned for the ProtoDUNE detectors hosted at the NA Neutrino Platform~\cite{charitonidis2017,vle:commissioning,protodune:tdr,protodune:results}. The low particle rate provided by the SPS slow extraction allows precise monitoring of the beam parameters, as, for instance, particle by particle momentum measurement and beam composition~(Sec.~\ref{sec:hadron_beamlines}). 
Before 1998, a different implementation of the SPS slow extraction was used to extract the beam in a few hundred \textmu s up to a few ms long spills, serving the West Area (WA) neutrino facility~\cite{wa:nbb,chds:crosssec,chds:theta,wanf_horn}, decommissioned in 2004. This type of extraction, referred to as ``fast-resonant'' or ``fast-slow''~\cite{sps:fastslow}, was based on a half-integer resonance and fast-discharge quadrupoles.
Recent slow extraction studies have been carried out at the SPS in the context of the ENUBET project~\cite{ENUBET_proposal}, aimed at delivering the technology of monitored neutrino beams. In ENUBET, the pile-up at the instrumented decay volume poses stringent limits on the particle rate, making the use of a $\mathcal{O}(\text{s})$-long slow extraction the preferred option.
However, to retain the possibility of using a magnetic horn or reduce the cosmic background at the neutrino detector, a new pulsed slow extraction scheme has been developed and tested. It consists in the extraction of $2$-to-$10$~ms proton pulses, repeated at $10$~Hz for the full extraction time~\cite{pari:tesi}. An example of this new scheme, the {\em burst-mode slow extraction}, can be seen in Fig.~\ref{fig:burstspill}.
The experimental results of the study have shown that this extraction method can be readily implemented and optimized at the SPS, based on the nominal slow extraction settings. During the machine development studies performed before the CERN Long Shutdown 2 (LS2, 2019-2021), the minimum observed temporal length of the pulses was $10$~ms. Dedicated simulation studies performed later on showed the possibility to span a $2$-$10$~ms range, and they will be validated after LS2. 

Unlike fast extractions, the slow extraction is an intrinsically lossy scheme: as the unstable particles gradually spiral out of the ring through the electrostatic septum, a small fraction of them will hit the septum blade. In the SPS, another fraction of the beam is lost on the downstream septa used to split the beam among the different transfer lines. As a consequence, the maximum number of integrated protons-on-target (POT) is limited by the activation of the accelerator equipment and the radiation hazard due to dose exposure of the workers. The number of POT extracted at the SPS in $2018$ of about $1.2\times10^{19}$, which represents a record since the WA times, is still far from the typical yearly POT of a fast-extracted neutrino beam. Even if significantly higher yearly rates are within the machine capabilities, important reductions of the extraction and splitting losses will be required to reach them. A dedicated campaign 
is ongoing at the SPS to reach this goal~\cite{slawg:2017,slawg:2019,matt:sx2019}.
For instance, local shadowing techniques as the insertion of a passive diffuser or a silicon bent crystal in front of the electrostatic septum have shown, during machine tests, loss reductions of about $10$-$15\%$ and $20$-$40\%$, respectively~\cite{brennan:diffuser,francesco:crystal,luigi:crystal}. Other techniques based on machine optics have also been investigated. In particular, first tests of octupole-based separatrix folding have shown a loss reduction of $\sim 40\%$ at the electrostatic septum~\cite{matt:octu}.
Long-standing efforts of reduction of the slow extraction losses are being undertaken at a global scale (e.g.~\cite{masahito-san:dynamic,vlad:octu}), calling for exciting developments in the future years.

The quality of the extracted spill is another critical issue of slow extraction: for precision experiments, a deviation of the spill from a constant particle flux might represent an issue (e.g.~higher counting uncertainties and pile-up). 
To remove the high-frequency components of the RF, the beam is often de-bunched. In this case, the dominant frequencies that survive in the extracted spill are typically the power supply harmonics and other noise coming from the magnet currents. Mitigation techniques to dump these parasitic frequencies 
have been recently developed at GSI~\cite{gsi:ripples}, J-PARC~\cite{jparc:ripples}, and the SPS~\cite{pari:tesi}.\\

To monitor the primary protons of a neutrino beam experiment, dedicated diagnostic devices are installed in the proton transfer line before the hadron production target. 
This solution typically constrains the total neutrino flux uncertainty derived from the primary proton beam to a maximum of $1$-$2$\%~\cite{miniboone:flux,t2k:flux,numi:flux}, which, in standard neutrino beamlines, represents a sub-leading contribution with respect to the dominant hadron production uncertainties.
A great variety of diagnostic devices is available for particle accelerators~\cite{forck:diag,cas:diag,strehl:binstr}, according to their main energy, intensity, extraction scheme, particle type, etc. The most relevant primary diagnostic instruments for neutrino beams are briefly described in the following.
\begin{itemize}
\item The total proton intensity is typically measured with Beam Current Transformers (BCT), non-destructive inductive devices that provide an absolute measurement of the total beam charge. Many variants are available for different types of beams, as their performance is strongly dependent on the spill length and instantaneous beam current~\cite{bct:tutorial,bct:original,sps:dcbct,lhc:fbct}.

\item Another important class of devices is the Secondary Emission Monitors (SEM). They can be used for different applications and consist of thin, metallic foils that are crossed by the beam to measure the emitted charge. When their signal is time-sampled, they can be used to measure the beam intensity profile during the extraction (Fig.~\ref{fig:burstspill}). SEM that are made up of several ribbons can be used to measure the transverse beam profile.
These devices can provide a measurement of the absolute beam intensity, too. However, their absolute calibration is prone to subtle systematic effects like charge emission due to activation~\cite{sps:foils}. SEM are particularly useful with $\mathcal{O}(\text{s})$-long slow extracted spills, which are difficult to measure with a BCT. Many recent developments have shown performance improvements and possible alternatives, as, for instance, the graphite-based profile monitor at J-PARC~\cite{jparc:gsem} and the Cherenkov detector for proton Flux Measurments (CpFM) tested at the SPS~\cite{puill:cpfm,francesca:cpfm,francesca:tesi}.
Other commonly used devices for beam profile measurements are the Optical Transition Radiation (OTR) screens. These are particularly convenient for ultra-relativistic beams, where the particles crossing a thin foil emit transition radiation, due to the change of medium, that can be recorded by light detectors.

\item Beam Position Monitors (BPM) are a class of devices used to measure the transverse position of the beam inside the vacuum chamber or in air~\cite{cngs:bpm}. They are non-destructive devices composed of a set of two or four electrodes, which act through 
capacitive pick-up. Because of this, they require short pulses or bunched beams: they are often replaced by SEM or OTR screens in slow-extracted beams~\cite{sps:foils}, or paired to them for increased reliability~\cite{malika:cngs}.

\item Beam Loss Monitors (BLM) are crucial machine-protection elements. They measure beam losses, i.e.~showers originated from the beam hitting or scraping a beamline element, and provide feedback to the control system. BLM are available in several types, but the most common in high-energy machines are based on ionization chambers~\cite{lhc:blm} or proportional counters because of radiation hardness constraints. In slow extractions, BLM are the ideal tool to evaluate the effectiveness of loss-reduction techniques.
\end{itemize}

These  devices are standard tools in any neutrino beam~\cite{t2k:overview,numi:beam,miniboone:flux,wanf:muflux,wanf:align,malika:cngs,cngs:instr} and represent the state-of-the-art in the diagnostics of primary protons. Typically, a few BCTs (included the ones mounted on the ring) yield the most accurate measurement of the beam intensity. BPM are used to monitor the position of the beam and verify its alignment with the target. SEM and OTR screens are used to measure the beam profile and validate the calculated optical beam parameters (e.g.~Twiss functions and emittance). They also monitor the beam position and alignment with respect to the target. 
A comprehensive report on high-intensity detectors, suitable for measuring the intensity of the primary beam on the secondary target can be found in Ref.~\cite{RhodriRep}.

\section{Target}
\label{sec:target}
Secondary particle productions hold the key for neutrino experiments, as the neutrino energy spectrum, flux, and kinematic distributions are strictly related to the secondaries emitted at the target. Therefore, it is fundamental for future experiments to have a precise knowledge of the interactions of the primary beam with the secondary target. 
As discussed in the previous sections, conventional neutrino beams are obtained from the decays of charged mesons ($\pi$, K) produced by protons impinging on a target. High-intensity neutrino beams -- often called {\em Superbeams} -- produced by multi-MW proton accelerators rely extensively on low-Z targets~\cite{Simos2012}. The higher beam power on the target compensates for the reduced yield compared to high-Z metallic compounds. Specifically, beryllium, graphite, or other carbon-based compounds, nickel-chromium superalloys, and mercury are the most common choices, given their robustness to radiation damage~\cite{Simos2019}. Granular or fragmented metallic targets have been also considered given their ability to absorb higher beam power, with promising results ~\cite{Caretta12, Caretta15}.

Neutrino experiments need to extrapolate and determine the flux of neutrinos from the yield of secondary hadrons produced by the target. This is often the most important source of systematic uncertainties. While it is possible to derive models for secondary production by fitting experimental data~\cite{Bonesini2001}, the uncertainties remain high ($>10$\%), especially for the low energy region, as reported in several surveys~\cite{Chemakin2008,Abgrall2016,Catanesi2008}. 
The number of $\pi^{+}$ produced per POT grows linearly with the primary proton momenta~\cite{Feynman1969}, which determines the accessible momenta for the secondaries and, hence, the neutrinos. The neutrino beam energy range is set by focusing only pions and kaons around a given  central momentum. While lower energy proton machines such as J-PARC MR~\cite{JPARC} can be operated at a higher repetition rate, the secondary yield is suppressed linearly with the proton energy compared, for instance, to the CERN SPS that can run  with $4\times10^{13}$ $400$~GeV/c protons extracted in $4.8$ seconds (lower energy favouring longer spill). Therefore, neutrino beam designers tune several parameters of the primary proton source, e.g.~the primary proton momentum, extracted intensity, and repetition rate, to reach the sought-for intensity of the facility. The target optimization plays the most important role in the optimization of secondary yields. 
Besides the choice of target materials, the target geometry determines the re-interaction probability and absorption of the secondary particles coming out of the target after the collision with the primaries. The optimisation of the target geometry, however, is not only driven by the produced yields, but also by mechanical constraints and cooling requirements, especially in the cases of high-intensity beams~\cite{Bruno2011}. In CNGS, for instance, the target was  an assembly of air-cooled, thin graphite rods, acting as multiple targets that could be hit separately. In this way, the spare targets were already located on-site.

Heat management is another parameter that needs consideration, especially in Superbeams. The energy deposited in the target material by the primary beam results in tensile stresses radiating out from the central beam spot~\cite{Hurh2012}. The main parameters that determine the heat endurance are the specific heat and the tensile properties of the medium. These properties drive the choice of the heat dissipation system. 
Heat dissipation may occur through the container that holds the target and the heat transfer from the target material can be either conductive or radiant in solid targets. Conductive internal heat transfer is very effective in metal targets, while radiant heat transfer is better suited for high-emissivity refractory materials and inserts, such as graphite~\cite{Popescu2020}. 

The target's overall size plays a central role in the cooling design. An increased surface implies a larger radiative heat transfer and the change of the size of a target is driven by the absorption/emission parameters of the material. 
Increasing the size could lead to a lower production yield due to secondary re-interactions inside the target. The optimization is then a trade-off between mechanical robustness against heating and the effective interaction length crossed by the secondaries.

An interesting example is an optimization performed for monitored neutrino beams and, hence, slowly extracted beams, by ENUBET. The team has conducted extensive optimization studies based on the FLUKA~\cite{FLUKA1,FLUKA2} and G4beamline~\cite{g4beamline} 
simulation codes, using graphite (density $2.2$~g/cm$^{3}$), beryllium (density $1.81$~g/cm$^{3}$), Inconel (density $8.2$~g/cm$^{3}$) and some high-Z materials. Each target prototype is modeled geometrically as a cylinder with variable radii between $10$ and $30$~mm and various lengths extending from $5$ to $140$~cm. 
\begin{figure}[h]
\centering
\includegraphics[scale=0.3]{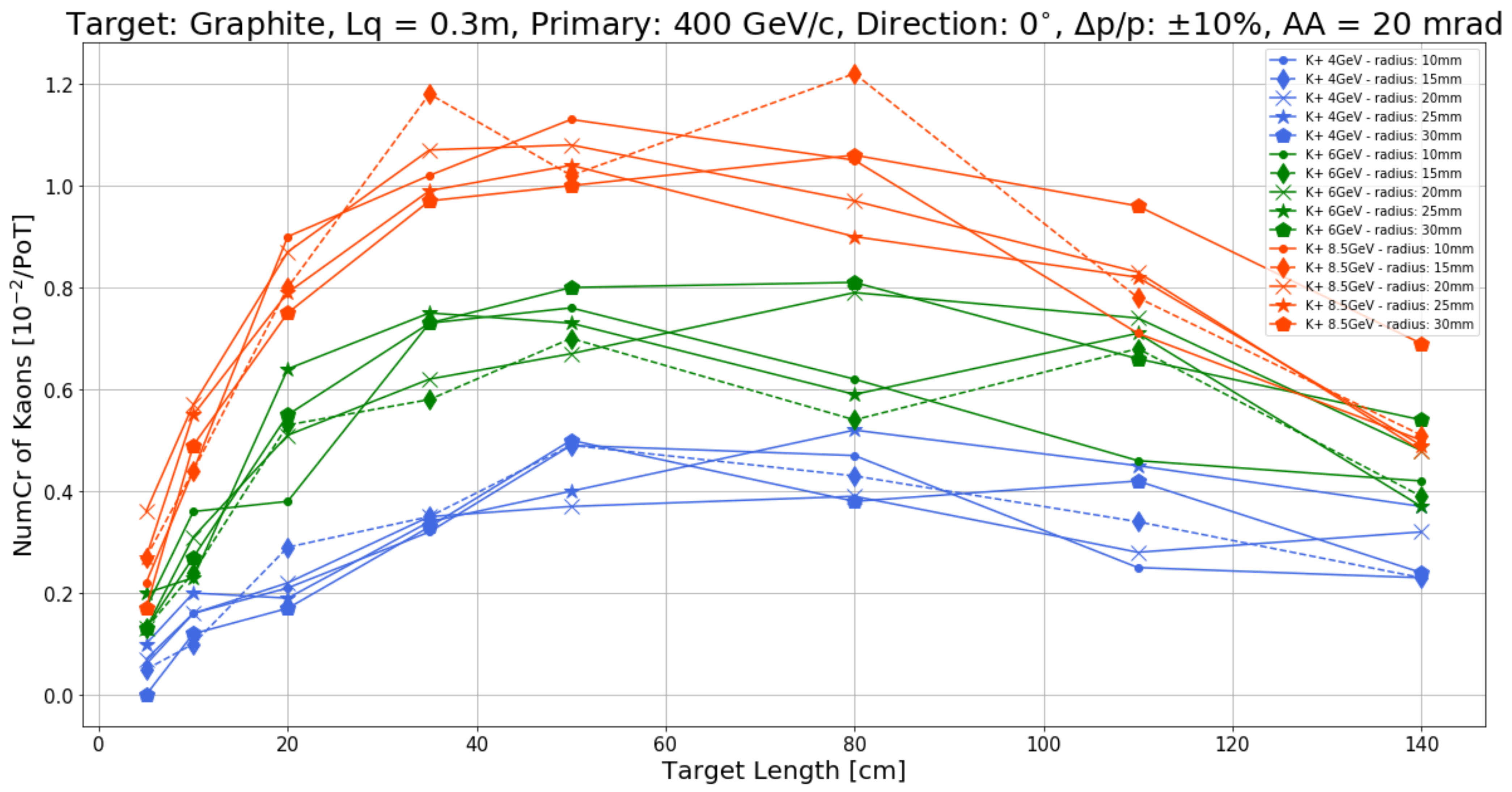}
    \caption{Kaon yields as a function of the graphite target length. The primary beam simulated is a $400$~GeV/c proton beam. The figure of merit for this study is the number of kaons of given energy with $10\%$ momentum bite that enters an ideal beamline with $\pm 20$~mrad angular acceptance (AA) in both planes, placed $30\,cm$ after the target (Lq). The error bars are not plotted to ease the reading; statistical errors are negligible ($1$\%), while the Monte-Carlo systematics amounts to $\sim 20$\%.} 
    \label{fig:K_prod}
\end{figure}
The target optimization was based on the production yield of $K^{+}$ particles into a typical secondary beamline acceptance ($\sim 20$~mrad)  and the intrinsic material characteristics (specific heat and density). 
The kaon yields are depicted in Fig.~\ref{fig:K_prod}. Eventually, the most promising materials for monitored neutrino beams are graphite and Inconel-718. 
Graphite is a known and well-tested material employed in several neutrino beams thanks to its heat endurance and production yields~\cite{Hurh2012}; Inconel is quite a novel choice that is under consideration for nuSTORM~\cite{Adey:2013pio} and ENUBET, but already at use at CERN in other applications (like the new CERN-PS East Area Beam Stoppers).

The main cooling technique~\cite{Popescu2020} for targets in Superbeams remains the water-cooling circuit. Heat is transferred to the water through an intermediate step either consisting of solid legs/fins, a gas mixture, or a heat pipe since the high temperature reached by the targets and the low boiling point of water do not allow these elements to be put in direct contact. Water cooling is employed in the NuMI beam serving the \nova~\cite{Ayres:2004js} experiment at Fermilab, where a $1.2$~m long segmented target made of graphite is cooled by  properly spaced fins and water circulating in a base-plate at the bottom of the fins~\cite{Jyoti2017}.

\section{Hadron yields}
\label{sec:hadron_yields}

Nowadays, hadron production experiments are critical for high precision accelerator neutrino beams. The differential distributions predicted by Monte Carlo codes used to compute the hadron production
yields from fragmentation models are usually not able to reproduce the kinematic distribution of secondaries, even if the models are tuned with external data. Mastering the hadron production mechanisms at a level sufficient to predict the accepted flux before the focusing and decay volume in high-precision neutrino experiments is still outside the reach of these models. 

There are two types of hadron production experiments that are of relevance in our field:

\begin{itemize}
    \item General purpose experiments aimed at
    measuring differential cross sections for pion and kaon production using thin targets of various materials. The materials are chosen among the most common candidates for neutrino targets (see Sec.~\ref{sec:target})  
    \item Replica target experiments, where the yields and differential cross-sections are measured for an exact copy of the target used in the corresponding neutrino experiment.
\end{itemize}

\begin{figure}[h]
\centering
\includegraphics[scale=0.3]{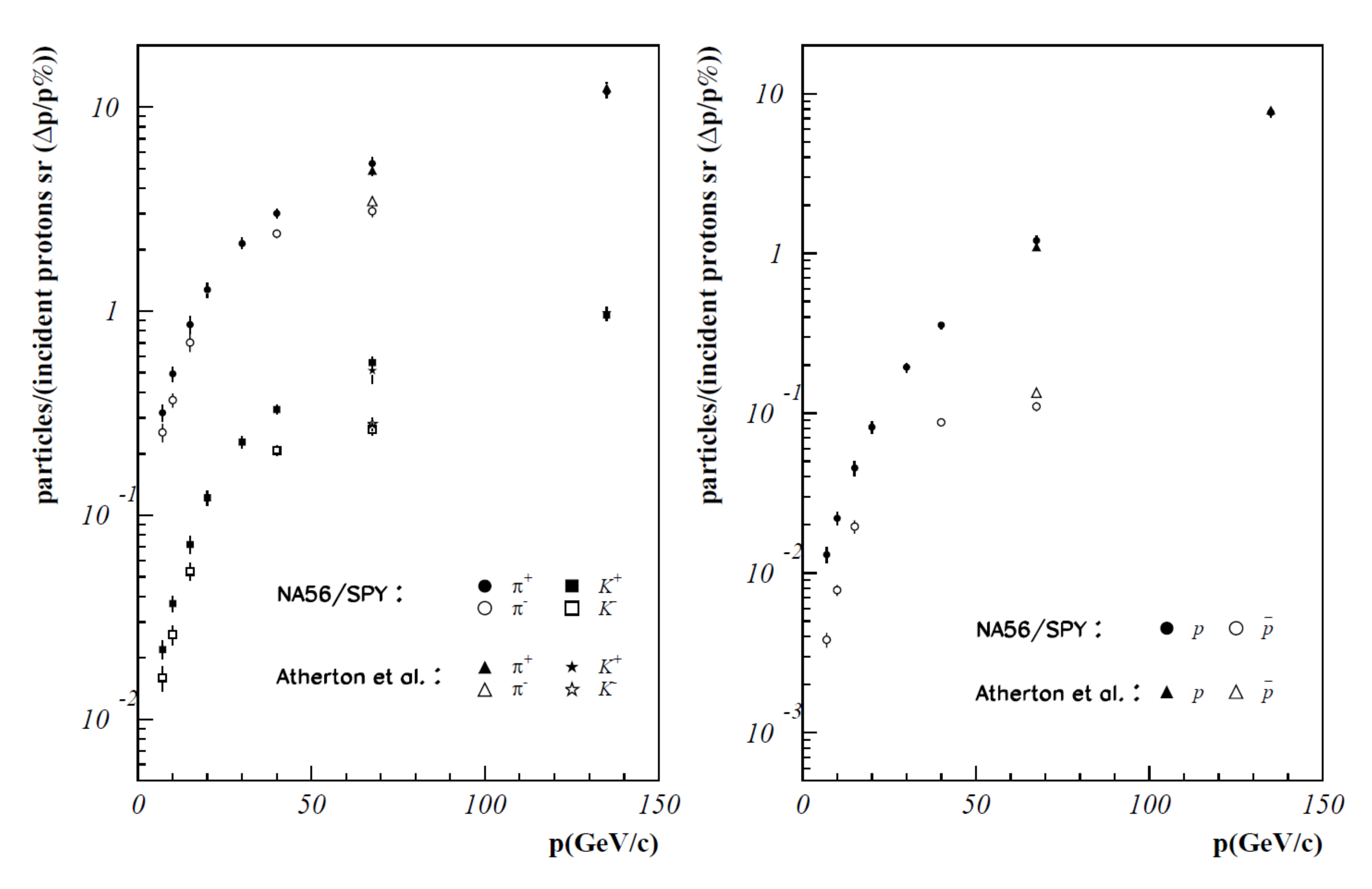}
    \caption{Pion and kaon yields from NA56/SPY. Reprinted by permission from Springer: \cite{Ambrosini:1999id}. Copyright: 1999.}
    \label{fig:spy_data}
\end{figure}

\noindent
In practice, both the knowledge of the cross-sections and the yields (typically performed with replica targets) is crucial for the correct normalization of the number of hadrons and, hence, neutrinos. 
The experiments performed in the 80s and 90s were mostly mass spectrometers with a narrow acceptance and excellent particle identification~\cite{Dore:2018ldz}. This choice was driven by the fact that the experiments were not designed as hadron production facilities but were based on existing experiments studying standard or exotic QCD processes. A renowned example is NA56/SPY~\cite{Ambrosini:1999id} at CERN that took data in 2002-2003 modifying NA52: a high precision mass spectrometer for the searches of strangelets in ion-ion interactions. Despite these limitations, SPY produced essential results for the design of the CERN-to Gran Sasso (CNGS) neutrino beam and the interpretation of the NOMAD data~\cite{Astier:2003gs}, both based on primary protons at $450$~GeV (the CERN SPS). The most precise data were taken on Beryllium in a secondary momentum range from $7$ to $135$~GeV/c and $p_T$ values up to $600$~MeV, where a precision of $5$-$10$\% was achieved. The $K^+/\pi^+$ ratio was measured in this momentum range with a typical precision of 3\%. Even if the precision achieved was quite high, the limited $p_T$ acceptance and the choice of thin beryllium targets increased the systematic budget when applied to real neutrino experiments due to the extrapolation to thick targets of different materials. SPY was, hence, useful in the design phase of CNGS to estimate the expected neutrino rates but did not constrain the actual neutrino flux below $25$\%. 

Modern hadron production experiments designed after the discovery of neutrino oscillations are dedicated facilities with a large $p_T$ acceptance, running both with thin and replica targets. HARP~\cite{Catanesi:2007ig} was the first dedicated hadron production experiment designed for a future Neutrino Factory and operated at the CERN PS (primary proton momentum: $25$~GeV/c) with a large acceptance spectrometer. The secondary momentum range was $0.5$-$8$~GeV/c and the maximum angle reached $250$~mrad. A special configuration (``large angle'') studied secondaries below $1$~GeV/c up to polar angles of $2150$~mrad. Even if the Neutrino Factory is still beyond the reach of current technologies, the HARP data were included to constrain the simulation of practically all low energy neutrino experiments: K2K, MiniBoone and, more recently, in the SBN Fermilab programme to search for sterile neutrinos~\cite{Antonello:2015lea}. For higher energies as the ones exploited by the NuMI beam, Fermilab devised a dedicated hadron production experiment called MIPP~\cite{Raja:2005kow} that reproduced the running conditions of \nova, MINOS and MINER$\nu$A. 
MIPP ran with the NuMI replica target (primary protons: $120$~GeV/c) accessing secondaries between $300$~MeV/c to $80$~GeV/c with a transverse momentum up to $2$~GeV/c. 

The broader purpose fixed-target experiment in use today is the NA61/SHINE~\cite{Abgrall:2014xwa} experiment at CERN. NA61 has a richer programme than its predecessors (NA52-NA56), but also an unprecedented angular acceptance. It is made up of 5 TPCs  -- including a couple of TPCs located in two special superconducting ``VERTEX'' magnets --  and a time-of-flight detector. NA61/SHINE measured the yields and the production cross-sections for T2K experiment ($31$~GeV/c proton on graphite targets) using both thin and replica targets. 
SHINE reported uncertainties in the production and inelastic cross-sections of the order of $O(2\text{-}8\%)$ for p+C, p+Be and p+Al at $60$ and $120$~GeV/c (see e.g.~\cite{SHINE2019}).
Before the CERN LS2, it also performed additional measurements for NuMI and recently extended beyond LS2 to serve DUNE, Hyper-Kamiokande and, possibly, NP06/ENUBET.  A special, low energy beam for measurements in the range of $1$-$10$~GeV/c is being currently considered by the collaboration for construction until 2027. 
The MIPP and NA61/SHINE data contributed to the world best measurement of the flux published by the MINER$\nu$A Collaboration and discussed in Sec.~\ref{sec:conclusions}.

\section{Focusing of secondary mesons}
\label{sec:focusing}

As discussed in Sec.~\ref{sec:narrow_wide}, the total neutrino flux depends on the angle of the neutrino decay with respect to the original meson direction, which typically scales as $\sim 1/\gamma$. Therefore, we can see from the bare-target kinematics of low energy neutrino experiments that the neutrino flux toward the detector is strongly suppressed if the secondaries are not focused. Focusing of secondaries is generally achieved either by magnetic horns or by quadrupole multiplets. 

\subsection{Horns}
\label{subsec:horn}

\begin{figure}[htb!]
    \centering
    \includegraphics[width=0.9\textwidth]{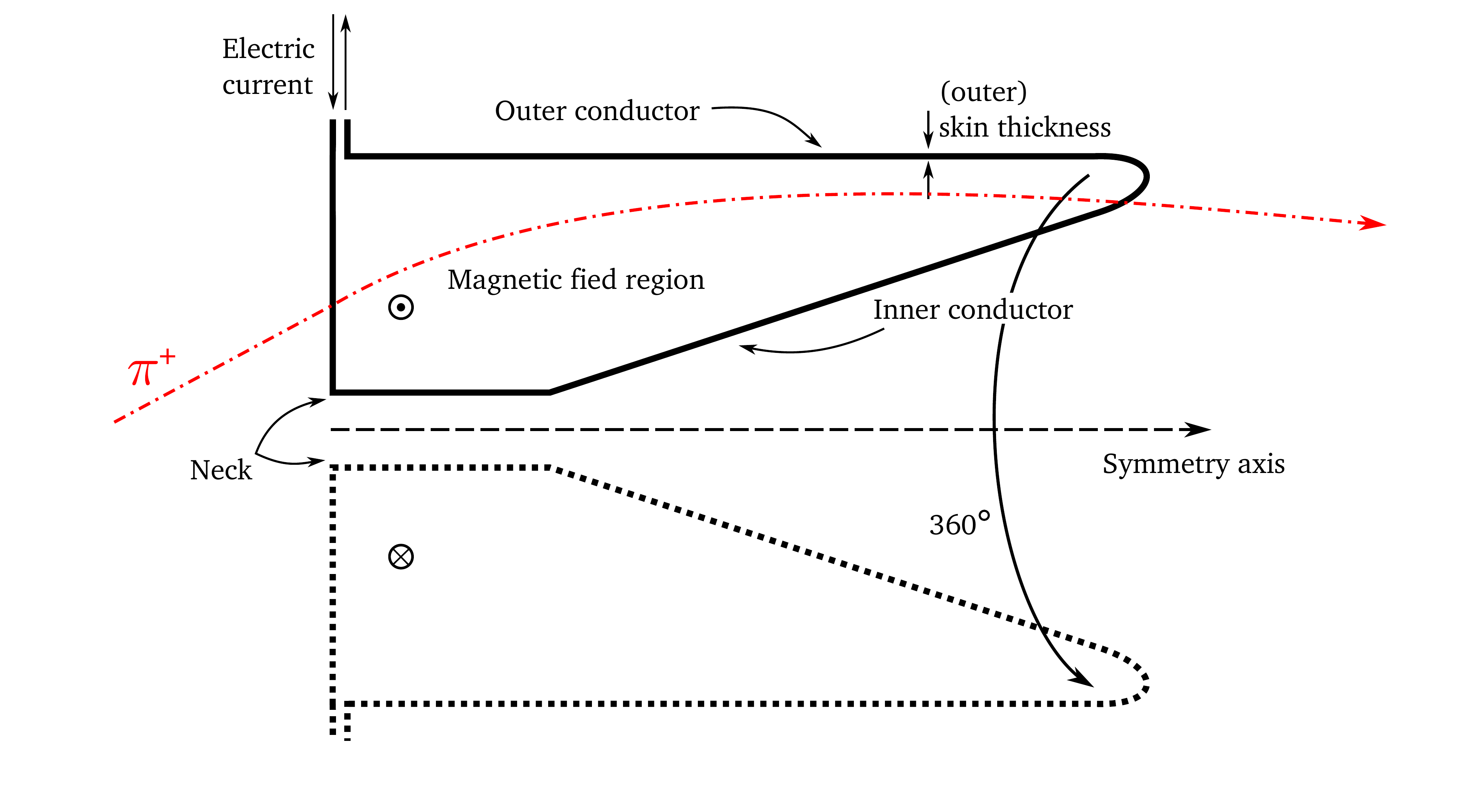}
    \caption{Schematic view of a magnetic horn: the device is axially symmetric and focuses particles of a given sign in both the transverse dimensions. The symmetry axis is also the target and beam axis.}
    \label{fig:horn}
\end{figure}

Magnetic horns were first proposed by S. van der Meer in 1961~\cite{HornMeer1961} and, since then, have been employed by all wide-band neutrino beams at the GeV scale.
A horn is composed of conductor sheets on which a current flows in the longitudinal direction during the proton extraction, producing a strong magnetic field.
A schematic representation of the device is reported in Fig.~\ref{fig:horn}.

The toroidal magnetic field of a horn provides a two-plane strong focusing for particles of a given sign and defocuses particles of opposite sign. Therefore, the focusing is accompanied by the suppression of the background from wrong-sign hadrons. 
Reverting the direction of flow of the current is equivalent to enriching the beam of
anti-neutrinos instead of neutrinos (or vice-versa). This is an essential requirement for long-baseline experiments, like DUNE or HK, where CP violation is established by the comparison of $\numunue$ and $\nubarmunubare$ transitions.

Various types of horns (e.g.~conical, parabolic, ellipsoidal, plugged) have been developed and employed over the years, and a comprehensive discussion can be found in~\cite{kopp2006}. 
All the future wide-band neutrino beams will rely on multiple consecutive horns (e.g.~two in HK, three in DUNE) to further increase the flux by re-capturing right-sign pions or kaons at the border of the acceptance of the previous horn and further charge-select and focus the secondary beam. 
Multi-horn systems have been also constructed and implemented at the CERN PS neutrino beam~\cite{Horn65CERN1, Horn65CERN2}, CERN SPS West Area Neutrino Facility (WANF), BNL neutrino program~\cite{ags:horns}, CNGS, and in the current generation of long-baseline experiments (T2K and \nova).

The magnetic field of a horn is proportional to the injected current and inversely proportional to the transverse radius; this is why very high currents ($>100$~kA) are typically used, placing the target in the proximity of the horn inner conductor and, often, integrated inside the neck.
The horn conductors undergo extremely large thermo-mechanical stresses. The pulsed currents of hundreds of kA induce a strong Joule heating on the device, which sums up to the thermal load coming from the energy deposited by the secondary particles crossing the conductor sheets. Besides, the very high magnetic fields generate strong forces, which might distort the horn shape. The increase of temperature also reduces the tensile strength, further weakening the horn structural stability. For this reason, a horn is always coupled to a complex cooling system, usually made by an extensive network of pipes and nozzles that spray water on the conductors. The power supply and the radiation damage of the connection cables are sources of potential failures, too, and a careful study of every part of the horn infrastructure is mandatory for stable operation~\cite{t2k:horn}.
On top of this, extended running can make the horn components extremely radioactive, imposing further limits on repairs and interventions. 
As a matter of fact, horns are among the most delicate parts of any beamline, and failures are fairly common~\cite{horn:opexp}.

The length, intensity, and repetition rate of the proton extraction are the most critical parameters for a horn design, operation, and lifetime, together with the value of the horn current. Typical configurations used for several years in various neutrino beamlines are reported in Table~\ref{tab:horn}, where a correlation between these parameters is quite evident. The lower the horn peak current, pulse length, proton intensity and energy, the higher the repetition rate achievable for stable operation. Hence, a horn operated with a slow extraction like the one of the WANF ($6$ ms proton pulses) required a low current and repetition rate for a stable operation.
\begin{table}[htb!]
\caption{Summary of the main horn and extraction parameters for different neutrino beams.
The repetition rates of CNGS and WANF consisted in two separate proton pulses, separated by the time interval reported in the table. ``Pulse length'' is the time duration of the current pulse from the power supply, while ``Current`` refers to the peak current value of the pulse.}
\label{tab:horn}
\centering
\begin{tabular}{lcccccc}
\toprule
\textbf{Beamline} & \textbf{Current} & \textbf{Pulse length} & \textbf{Repetition} &\textbf{Intensity} & \textbf{Momentum} & \textbf{Extraction} \\
 & \textbf{(kA)} & \textbf{(ms)} & & \textbf{(protons)} & \textbf{(GeV/c)} \\
CNGS~\cite{cngs:ctd,horn:opexp,gilardoni:tesi} & $150$ & $3.8$ & $2$ by $50$~ms, $0.3$~Hz & $2.4\times10^{13}$ & $400$ & Fast \\
MiniBooNE~\cite{miniboone:flux}  & $170$ & $0.143$ & $5$ & $5\times10^{12}$ & $8.9$ & Fast\\
NuMI~\cite{numi:beam}  & $205$ & $2.3$ & $0.5$ & $3\times10^{13}$ & $120$ & Fast \\
T2K~\cite{t2k:horn,t2k:overview}  & $320$ & $2.4$ & $0.5$ & $3\times10^{14}$ & $30$ & Fast\\
WANF~\cite{wanf_horn,wanf:muflux}  & $100$ & $\gtrsim 7$ & $2$ by $2.7$~s, $0.07$~Hz & $2.4\times10^{13}$ &  $330$-$450$ & Slow ($6$~ms) \\
\bottomrule
\end{tabular}
\end{table}

Developing horns with significantly improved operational specifications than any of the reported configurations of Table~\ref{tab:horn} goes beyond the state-of-the-art, and requires dedicated studies or alternative solutions to the problem~\cite{horn:zimm}.
This has been the case for many proposed next-generation neutrino beams. For instance, a dedicated study of a $\sim 300$~kA horn with a $12.5$~Hz repetition rate had been carried out for the European Neutrino Superbeam (EURO$\nu$SB)~\cite{euronu:ov,euronu:hornps,gilardoni:tesi}. More recently, the European Spallation Source Neutrino Superbeam (ESS$\nu$SB)~\cite{essnusb_14,essnusb_19}, which would operate a proton extraction of $2.68$~ms pulses of $\sim10^{15}$ protons repeated at $14$~Hz, has chosen to develop a dedicated accumulator ring to compress the proton pulses to some \textmu s of temporal length~\cite{essnusb_accumulator_1,essnusb_accumulator_2}, in order to exploit standard horn technologies. The burst-mode slow extraction developed at the SPS as an option for the ENUBET monitored beam (Sec.~\ref{sec:proton}, Fig.~\ref{fig:burstspill}) also requires dedicated horn studies.

Aside from the hardware limitations and constraints, selecting the optimal horn geometry and implementation for a particular neutrino beam is a complex task. As the horn is a non-linear magnet (field $\propto~1/R$), its longitudinal profile plays an important role in the final focusing,
and -- by design -- particles need to cross its surface to be focused. Extensive numerical simulations based on particle tracking and interaction codes (e.g.~GEANT4, FLUKA, MARS~\cite{mars:ug}, G4beamline) are mandatory to quantify the focusing effectiveness. Each horn geometry has a quite high number of degrees of freedom (typically higher than $10$) which play a role in its performance, and numerous constraints coming from the hardware side. For this reason, horn designers perform multi-parameter numerical optimizations to find the best configuration, recently also exploiting meta-heuristic global optimization methods (e.g.~genetic algorithms)~\cite{andrea:horn,ichikawa-san:horn,nustorm:horn,pari:tesi}.

Horn diagnostics is pivotal not only for safety and reliability reasons. Variations in the current flowing through the horn and permanent deformations of the conductor generate systematic biases in the beam focusing, which ultimately affects the neutrino flux. 
Deformations change the beam optics from the design one, thus reducing the flux directed toward the Far Detector and increasing the production of background tertiaries. This background originates in unfocused particles hitting the downstream beamline elements. 
The horn currents are typically measured with a precision of $1$\%, and give a very small contribution to the flux systematic budget. For experiments where the target is located next to the horn, the systematics are driven by uncertainties in the relative target-to-horn distance~\cite{numi:flux,t2k:flux}. Their variations over time are difficult to be monitored because of the very harsh radiation environment, but their contributions are important mainly at the edge of the acceptance region and introduce systematics of the order of $2$-$5$\% only in limited regions of the neutrino spectrum.

As magnetic horns are characterized by a broad momentum acceptance, they are less effective for narrow-band neutrino beams, where an additional momentum selection stage based on radiation-hard magnetic elements (dipoles and quadrupoles) is required.

\subsection{Quadrupole multiplets}
\label{sec:quadrupole_multiplets}

An alternative to magnetic horns is quadrupole multiplets, forming an ``acceptance stage'' for the hadrons just downstream of the production target. Differently from a horn, a single quadrupole focuses the particles along one axis and defocuses them on the other: a net focusing effect in both transverse dimensions can be achieved with a combination of them (from two quadrupoles onward). Three quadrupoles in series (i.e.~a quadrupole triplet) are a common choice as a focusing stage because they allow achieving similar focusing proprieties both in the vertical and horizontal plane. Using doublets or triplets, the produced secondaries are accepted and collimated. The particles are charge and momentum-selected by inserting bending dipoles and slits along the beamline, and re-focused towards the decay volume. 
Quadrupoles, dipoles and other standard accelerator magnets are referred to as {\em static focusing elements} because they all run at constant currents (or are pulsed for long times, $O(s)$) for the entire duration of the experiment. 

In general, the quadrupole multiplets have a smaller acceptance than the horns, except for $\mathcal{O}(100)$~GeV/c secondaries~\cite{Conrad:1997ne}. 
Still, they offer major advantages. There is no need to pulse them and, therefore, they can be used in combination with a $\mathcal{O}(\text{s})$-long slow extraction. Standard large-aperture quadrupole magnets with gradients of the order of $\sim 10$~T$/$m are significantly less expensive, simpler, and robust than horns. 
Even if horns are a must for long-baseline experiments, the flux requirements of short-baseline experiments can be fulfilled by well-designed static systems even in the low-energy regime~\cite{Carey1970}. They can operate in DC or, generally, low currents of the order of 1~kA, thus they do not produce a major thermal load. They are particularly suitable for monitored neutrino beams because they work in slow extraction operation mode and reduce the overall length of the beamline from the target to the decay volume. Such length determines the number of kaons whose lepton cannot be monitored because the parent decays before the tunnel instrumentation. 

\section{Hadron Beam Lines}
\label{sec:hadron_beamlines}
 
As discussed above, the produced mesons at a target need to be focused, momentum selected and transported towards the decay volume. The beam inside the decay volume needs to be parallel or with a very small divergence, for reasons of containment within the decay region. In this way, we  minimize the probability of the secondary hadrons to directly interact with the walls or the instrumentation of the decay volume, thus creating {\em tertiary reinteraction} background. 

\begin{figure}[hbt!]
\centering
\includegraphics[scale=0.35]{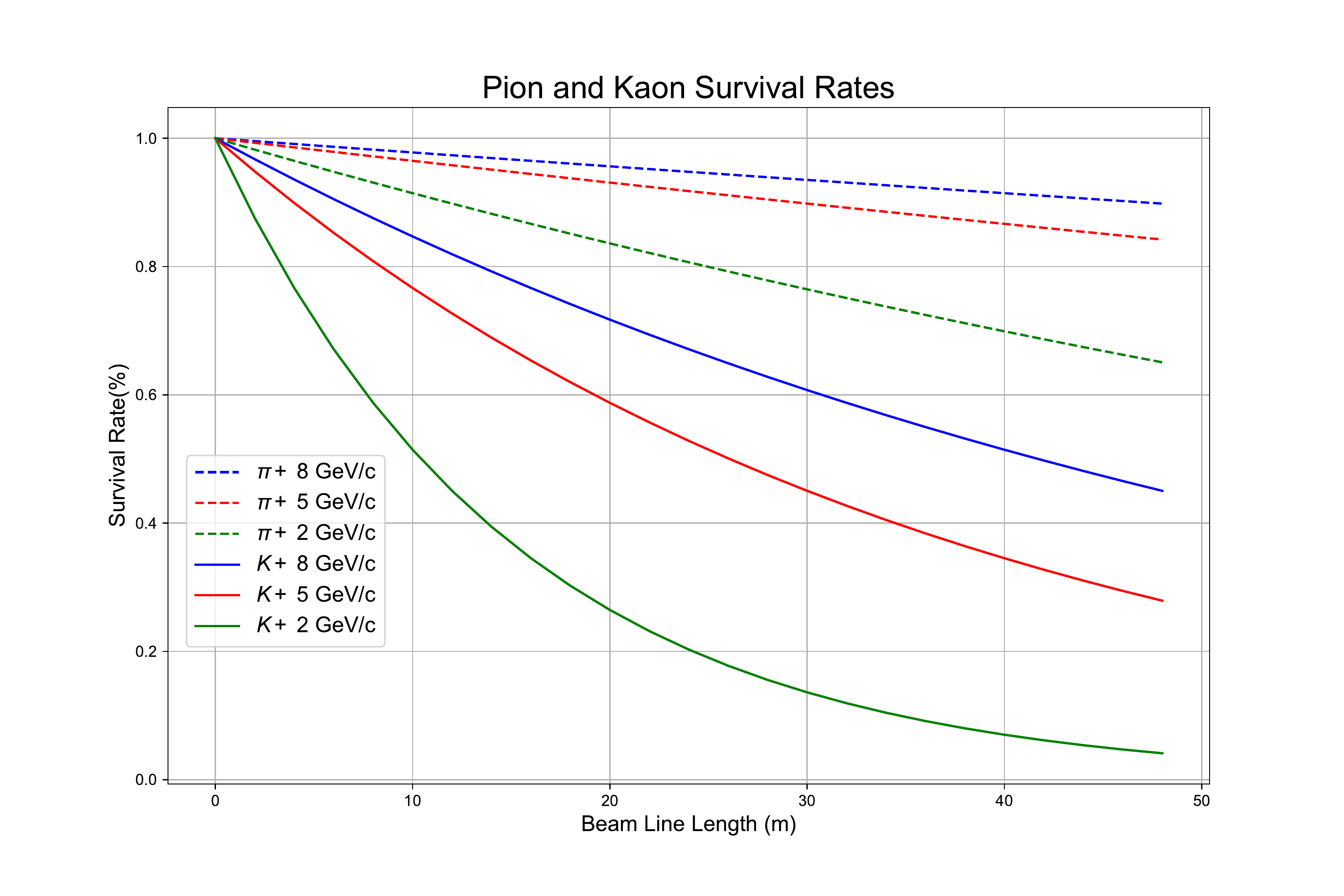}
    \caption{Survival probabilities for low momentum pions and kaons, assuming a beamline with a maximum length of $50$~m.}
    \label{fig:decays}
\end{figure}

A typical secondary beamline consists of three different "stages" constituting different elements. A schematic illustration of the basic stages is shown in Fig.~\ref{fig:narrow_wide}. All the designs of the hadron beamlines need to take into account the decay probabilities of the unstable mesons (typically pions and kaons) that should not decay before the decay volume. As an illustration, for low energy neutrino beams ($\leq 10$~GeV/c) the survival probabilities for pions and kaons for a beamline length of $50$~m is shown in Fig.~\ref{fig:decays}.

\subsection{Acceptance Stage}

In the case of conventional beams, a high-intensity, high-energy proton  beam with a small emittance impinges on a target material. The lower-energy secondaries are emitted in large production angles, and the non-interacting proton beam continues essentially un-deflected, but with an attenuated intensity that depends on the number of interaction lengths of the target. The first geometrical acceptance of the produced hadron phase-space is done in neutrino beams by the above-mentioned magnetic horns or quadrupole multiplets. Typically, quadrupoles may offer a final beamline acceptance of the order of a few tenths of milliradians in angle and tenths of mm in space. 
The positioning of the quadrupoles with respect to the target needs to be carefully optimized. In common practice, quadrupoles with large apertures, i.e radii -- $O(100 \mathrm{mm})$ -- are preferred, especially when the energy range of the selected secondaries is below $10$~GeV/c (see for example ref.~\cite{charitonidis2017}). 

\subsection{Momentum Selection \& Final Focusing Stages}
\label{subsec:momsel}

In low intensity charged particle beams,
the accepted particles from the first stage cross through a momentum-selection station,  consisting of a magnetic spectrometer. In most cases, it includes two or more dipole magnets, in combination with a quadrupole magnet and a collimator slit. The momentum selection comes with a (sometimes quite important) reduction of the overall rate. In case of slowly extracted or low rate beams, a momentum spectrometer made of tracking devices (such as fast wires or scintillating fiber monitors~\cite{XBPF}) can offer a particle-by-particle momentum reconstruction with a resolution down to $2.5$~\%, depending on the spatial resolution of the tracking devices employed. These techniques are discussed, for example, in Refs.~\cite{Booth2019, AtlasSPEC}. After the momentum selection stage, it is usual to recombine the dispersed rays so that the particles exiting the momentum selection have, at first-order, trajectories independent of their momentum within the accepted momentum band. For this reason,  we either implement a field-lens quadrupole or a ``geometrical'' recombination of the dispersion, depending on the exact configuration of the magnetic elements that generate the dispersion. 
After the momentum selection and recombination stages, a quadrupole multiplet (usually a doublet or triplet) is used to make the beam parallel towards the end of the transfer line (the decay volume in neutrino beams). The strength of this stage must be tuned according to the transverse and the longitudinal dimensions of the decay volume. In-between or downstream of these quadrupoles we place the particle identification instrumentation.

This class of designs can be applied also for narrow-band neutrino beams, adding the horn (Sec.~\ref{subsec:horn}) or working with radiation hard, large aperture magnets (Sec.~\ref{sec:quadrupole_multiplets}).
Given the large intensity, single-particle diagnostics is not a viable option and can be used only in low-rate calibration runs.

The transfer line between the target and the decay volume is highly tunable. 
Depending on the chosen target and the specific particle production, we can design a beamline that selects, accepts and transports a variety of momenta. This is usually achieved by scaling the strengths of the magnetic elements of the beam, i.e.~re-optimizing the currents. In this case, the particle identification instrumentation must be able to operate efficiently in a wide range of momenta. The power supplies used e.g., in the CERN North Area have an uncertainty of $\sim 0.15$~A and, for large deflection angles, may give uncertainties on the hadron momentum up to $\sim 3 \%$.

Clearly, high-intensity wide-band neutrino beams cannot afford a hadron, momentum-selective beamline due to the strong reduction of fluxes due to acceptance limitations. In these cases, the decay volume is located just after the horns as in Fig.~\ref{fig:narrow_wide} bottom.

\subsection {Particle Identification Instrumentation}

The particle identification instrumentation in the transfer line needs to take into account the hadrons rate, the background and the momentum range of the particles. In the case of low energy beams, the material budget imposed by the instrumentation needs to be taken into consideration, too. 

\begin{figure}[h]
\centering
\includegraphics[scale=0.4]{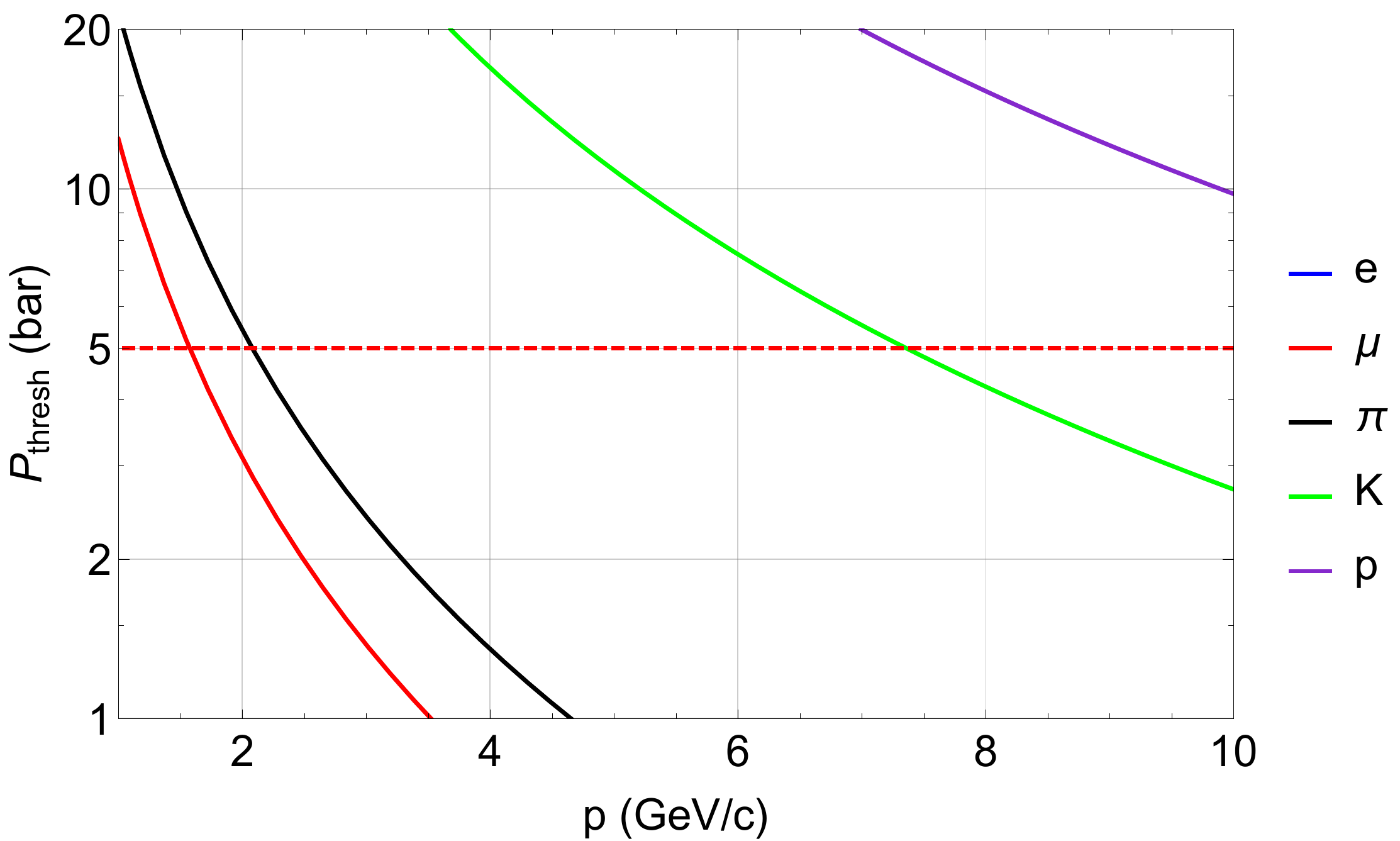}
    \caption{Threshold pressures of a threshold Cherenkov counter typically used in the CERN experimental areas. For energies above 4~GeV/c and using two counters, one at 5~bar and another e.g.~at 2~bar, protons can be separated by pions. Above 6~GeV/c kaons can be tagged, too. Electrons are always ultra-relativistic and can be identified at any momentum. }
    \label{fig:Cherenkov}
\end{figure}

For hadron momenta lower than $\leq 10$~GeV/c, particle identification can be achieved employing threshold Cherenkov counters and Time-Of-Flight (TOF) systems. The threshold pressure curves for a typical Cherenkov counter used at the CERN experimental areas filled with CO$_{2}$ can be seen in Fig.~\ref{fig:Cherenkov}. A standard technique employed at the CERN North Area to identify all charged particles above $4$~GeV/c makes use of two threshold Cherenkov counters, one filled with CO$_2$ and another one with R218 or R134a gases ~\cite{Charit_cherenkov, Charit_R134} and is described in Ref.~\cite{Charit_note}.

\begin{figure}[h]
\centering
\includegraphics[scale=0.4]{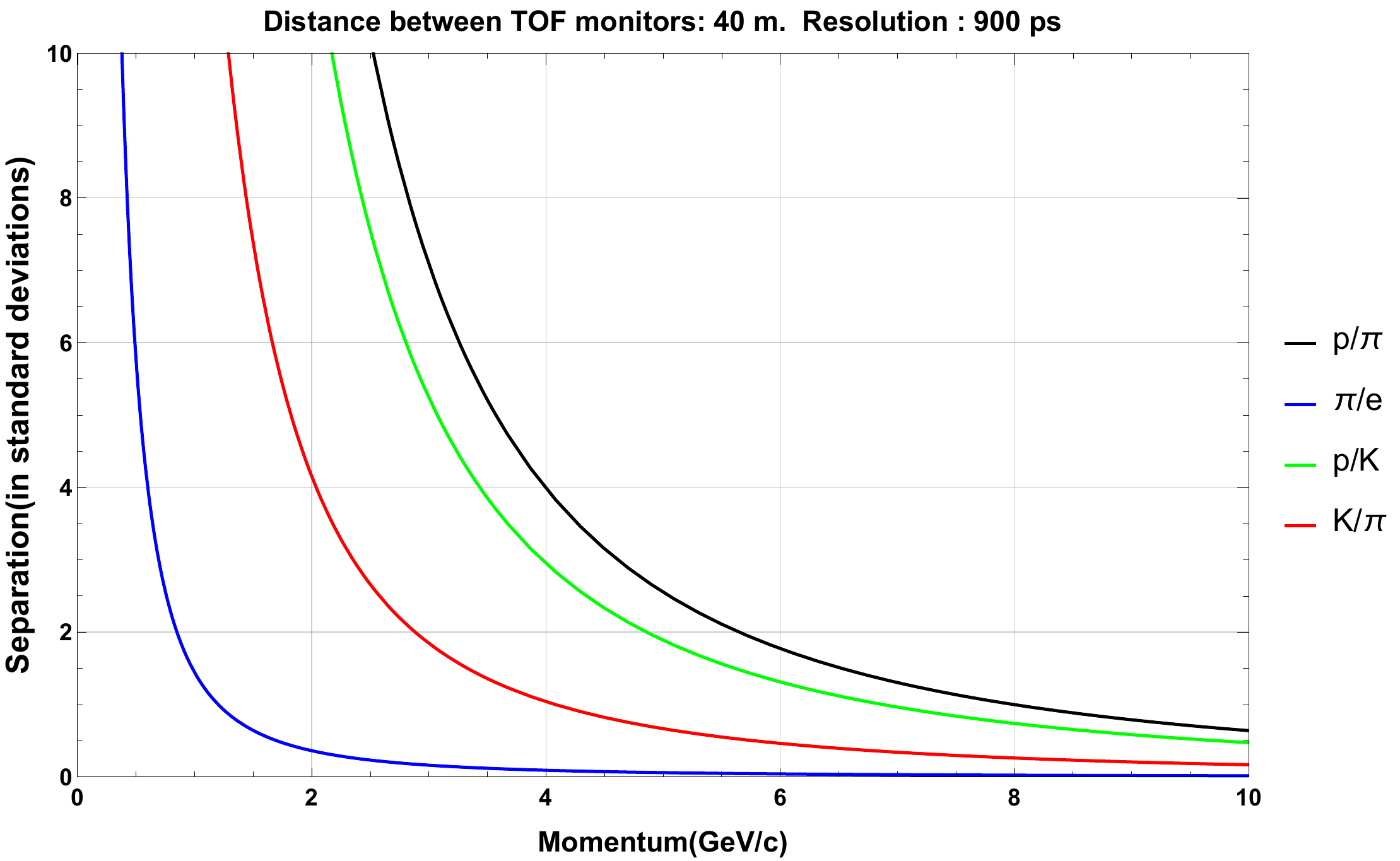}
    \caption{Separation of particles (expressed in standard deviations) for a  time-of-flight resolution of $900$~ps. Below $4$~GeV/c TOFs achieve an excellent separation  for protons, kaons and pions.} 
    \label{fig:ToF}
\end{figure}

For hadron momenta below 4~GeV/c, a TOF measurement ensures better performance. As an illustration, the separation capability among hadrons is shown in Fig.~\ref{fig:ToF}. Below 4~GeV/c, kaons are generally missing in the beam due to their short decay length, while the protons can be identified clearly from kaons and pions, and the separation power is inversely proportional to the time resolution of the system. 

In all cases above, the rate of the hadrons in the beamline and the material budget seen by the secondaries are crucial parameters. 
The aforementioned particle-identification (PID) techniques are effective for intensities significantly lower than $O(10^{11})$ particles per spill, limited by the pile-up in the scintillators as well as the TOF resolution. At the same time, these instruments contribute very little to the material budget -- about $0.2$ radiation lengths ($X_{0}$)  -- thus they do not contaminate or attenuate the hadron beam. Their use for reliable particle identification was sustainable in early neutrino beams, but these devices cannot stand the power of modern short and long-baseline facilities. However, even in very high intensity facilities, important information on the effectiveness of the focusing, the size of the background (e.g.~from interactions in the slits or the magnet's apertures), the momentum spread and the beam composition can be gained employing the instrumentation described above in dedicated low-intensity runs. 

\section{Decay Section}
\label{sec:decay_tunnel}

The length of the decay volume is generally optimized for a significant decay fraction of the pions. Since a pion travels about $56$~m per GeV, the length of the decay volumes is of the order of several hundred meters. A quite extreme example is CNGS, where the average neutrino energy was $17$~GeV. The facility thus required a 1~km long evacuated tunnel pointing to Gran Sasso (Italy). Due to the long baseline, the tilt of the tunnel was large: in this case, a possible ND would have been located hundreds of meters below the sea level, with a substantial increase of the cost. As demonstrated in Sec.~\ref{sec:ND}, the benefit of an ND for a pure $\nu_\tau$ appearance was marginal compared with the cost and the facility ran without an ND. Conversely, NuMI -- serving MINOS and \nova -- had a $675$~m long decay volume with a near detector located $900$~m after the target. DUNE has to face an issue similar to CNGS since the baseline is $2300$~km but an ND is mandatory for the precision measurement of CP violation. The challenge was solved by creating an artificial hill above the target station where the beam is steered down toward the Sanford Underground Research Facility (SURF), in South Dakota. In this case, the decay volume is just $194$~m long and the near detector is located in a relatively shallow hall $574$~m far from the target.  
This solution is feasible because the neutrino energy of DUNE ($\langle E_{\nu_\mu} \rangle \simeq 3$~GeV) is much smaller than CNGS ($17$~GeV). 

Several accelerator neutrino beams designed during the previous century used non-evacuated decay volumes. To prevent the interactions of secondaries with air, the volume was replaced by Helium, which has an interaction length of $426$~m. 
This solution is no more viable in contemporary physics because of the tremendous increase of the beam intensity, which produces tertiaries in helium and a dangerous air activation by tritium. Therefore, all tunnels are now evacuated at the mbar level or below using root pumps that can handle large volumes at a moderate cost. 

\subsection{The decay volume of monitored neutrino beams}

\begin{figure}[tb]
\centering
\includegraphics[scale=0.45]{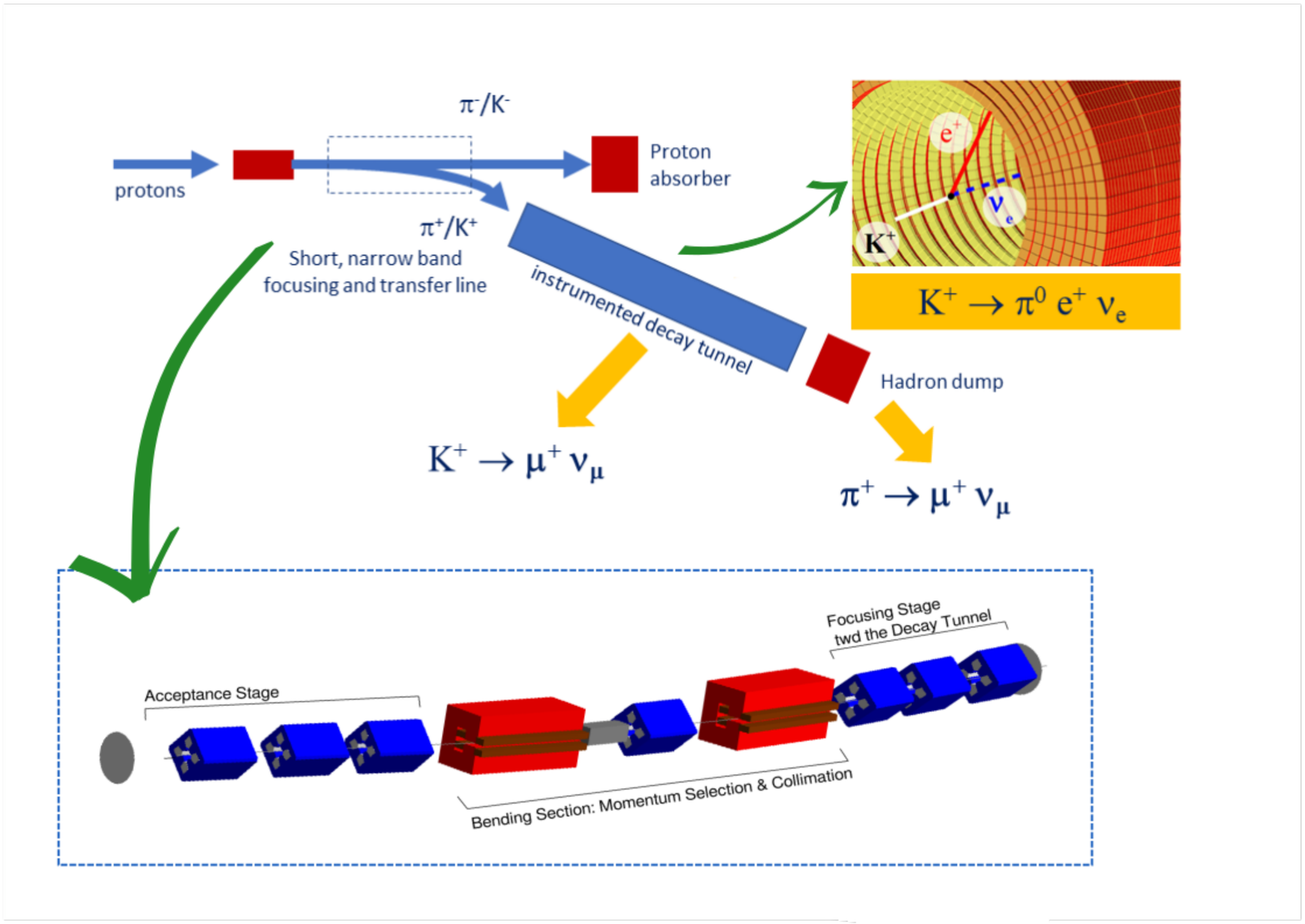}
    \caption{Schematics of the ENUBET facility. The decay volume instrumentation is shown in the top right inset; the (horn-less) focusing and transfer line is shown in the bottom inset. }
    \label{fig:beamlinescheme}
\end{figure}

The core device of a monitored neutrino beam is the instrumentation of the decay volume for the identification of the leptons (see Fig.~\ref{fig:beamlinescheme}). In the following, we will detail this instrumentation referring mainly to NP06/ENUBET, which has developed the most advanced design, to date.
In ENUBET, the vacuum requirement in the decay volume is slightly tighter ($\sim 0.1$~mbar) than standard neutrino beams, but can still be handled by root pumps. As mentioned above, monitored beams are neutrino beams equipped with radiation-hard instrumentation in the lateral walls of the decay volume to identify large-angle leptons produced by the decay of the secondaries. For ENUBET, the instrumentation identifies large-angle positrons from $K^+ \rightarrow e^+ \pi^0 \nu_e$ ($K_{e3}$) and muons from $K^+ \rightarrow \mu^+ \nu_\mu$ ($K_{\mu 2}$), as shown in Fig.~\ref{fig:beamlinescheme}. In this case, background reduction is essential. An interesting ancestor of ENUBET was the K2K~\cite{Ahn:2006zza} beamline, where a Cherenkov counter was positioned inside the tunnel in dedicated low-intensity runs to estimate the kaon contamination and, in general, the composition of the hadrons. Similarly, during the first attempts to build a ``tagged neutrino beam'' in USSR (see Sec.~\ref{sec:tagged}) physicists installed trackers to measure the trajectory and the time-of-flight of the secondaries~\cite{ammosov,bernstein}.
Other options resorted to transforming the tunnel itself in a Cherenkov radiator to identify the muons but the extremely harsh environment and the additional material in the tunnel makes the use of Cherenkov tagging quite challenging~\cite{Ludovici:1996sx}.
A completely different approach for the instrumentation of the channel was proposed in 2010~\cite{Ludovici:2010ci}: here, the authors gave up the possibility to intercept the core of the hadron beam and designed a detector to track only large-angle particles. This ``beam scratching'' approach was perfected in 2015, where a proposal to instrument just the walls of the tunnel was put forward~\cite{Longhin:2014yta}.

Unlike earlier attempts, the proposal of Ref.~\cite{Longhin:2014yta} reaps a decade-long R\&D for detectors capable of sustaining high particle rate in a harsh radiation environment. The current ENUBET design is based on cost-effective detectors to instrument a significant fraction ($>50$\%) of the decay volume. Since ENUBET is tuned to maximise the kaon decays instead of the pion decays, the tunnel is unusually short ($40$~m), which further reduces the cost of the instrumentation. The ENUBET positron detectors are a set of modules made of iron interleaved by plastic scintillator slabs, whose light is read by wavelength shifter fibers. The fibers transport the light on the outer part of the tunnel radius where the photosensors (Silicon Photomutipliers, SiPMs) are shielded against the expected neutron flux ($10^9$-$10^{10}$~n/cm$^2$). 
Each module samples $4.3$~$X_0$ and therefore monitors the development of electromagnetic and hadronic showers, together with the trajectory of the minimum-ionizing-particles (mips). $e/\gamma$ separation is achieved by a photon veto made of scintillator tiles located in the inner radius of the cylinder. Using three radial layers of modules, ENUBET achieved a signal-to-noise of 2 for positrons from $K_{e3}$ and 6 for muons from $K_{\mu2}$. These results were the outcome of a full GEANT4 \& FLUKA simulation and were validated by a beam test campaign carried on at CERN's East Area between 2017 and 2018~\cite{Acerbi:2020nwd}.

Two other paths explored by the ENUBET detector R\&D turned out to be of significant interest for applied physics. Firstly, in 2017 we developed a segmented calorimeter made of ``Ultra-Compact-Modules'' (UCMs~\cite{Berra:2017rsi}) where the SiPMs are embedded in the bulk of the calorimeter and the light is read by fibers punching both the iron and the scintillator tiles, perpendicularly to the tile plane (the so-called shashlik readout). This is a very elegant adaptation of shashlik calorimetry that solves the most important drawback of shashlik-based devices: the inability to build detectors with a fine longitudinal segmentation due to the presence of large dead areas. The compactness of the SiPMs reduces the dead zones to a few percent of the UCM module~\cite{Ballerini:2018hus}. This is a major step forward in shashlik calorimetry, which is employed in collider physics since the 80s.  This solution is usable also in a monitored neutrino beam but it has been replaced by the design mentioned above to have a $20\times$ safety margin on irradiation damage~\cite{Acerbi:2019wti}. In the shashlik case, the SiPMs are exposed to the neutrons produced in the core of the hadronic shower and the irradiation fluence increases substantially. 
Besides, ENUBET proved the possibility to use an extremely radiation-hard scintillator based on polysiloxane. This is the first application of such material in high-energy physics~\cite{Acerbi:2020itd} and is one of the options for the instrumentation of the muon trackers after the hadron dump, as described in Sec.~\ref{sec:hadron_dump}.

\subsection{Flux monitoring at the Neutrino Factories}

\begin{figure}[t]
\centering
\includegraphics[scale=0.4]{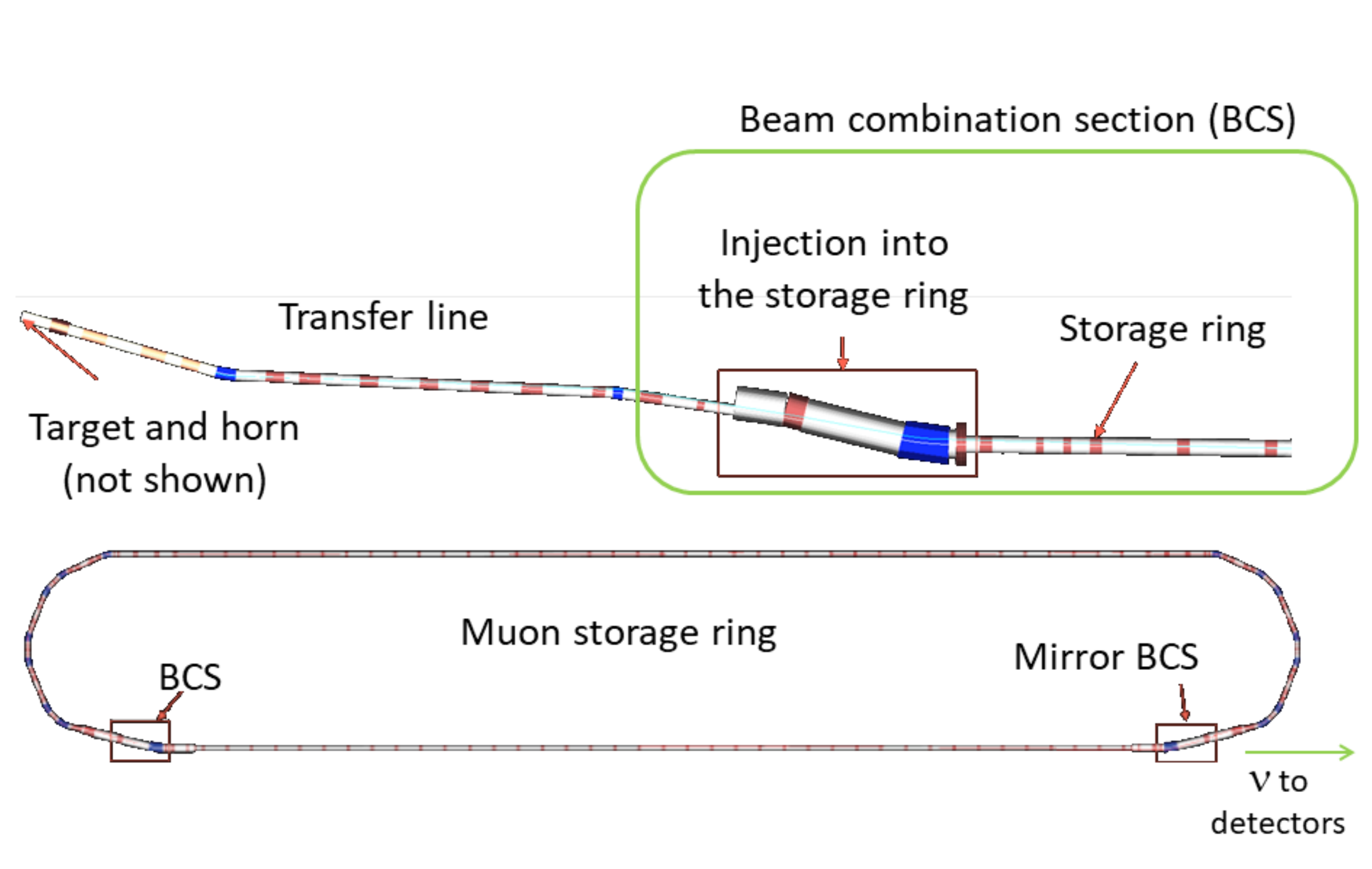}
    \caption{Schematics of the nuSTORM facility. (top) the pions focused upstream by a horn (not shown) are transported and injected into the storage ring. (bottom) the storage ring made of a large acceptance lattice with quadrupoles and bending dipoles. The muons from pion decays produce neutrinos of both electron and muon flavor mostly in the straight session. The Beam Current Transformer is located inside the beampipe of the storage ring.}
    \label{fig:schematics_nustorm}
\end{figure}

\begin{figure}[h]
\centering
\includegraphics[scale=0.4]{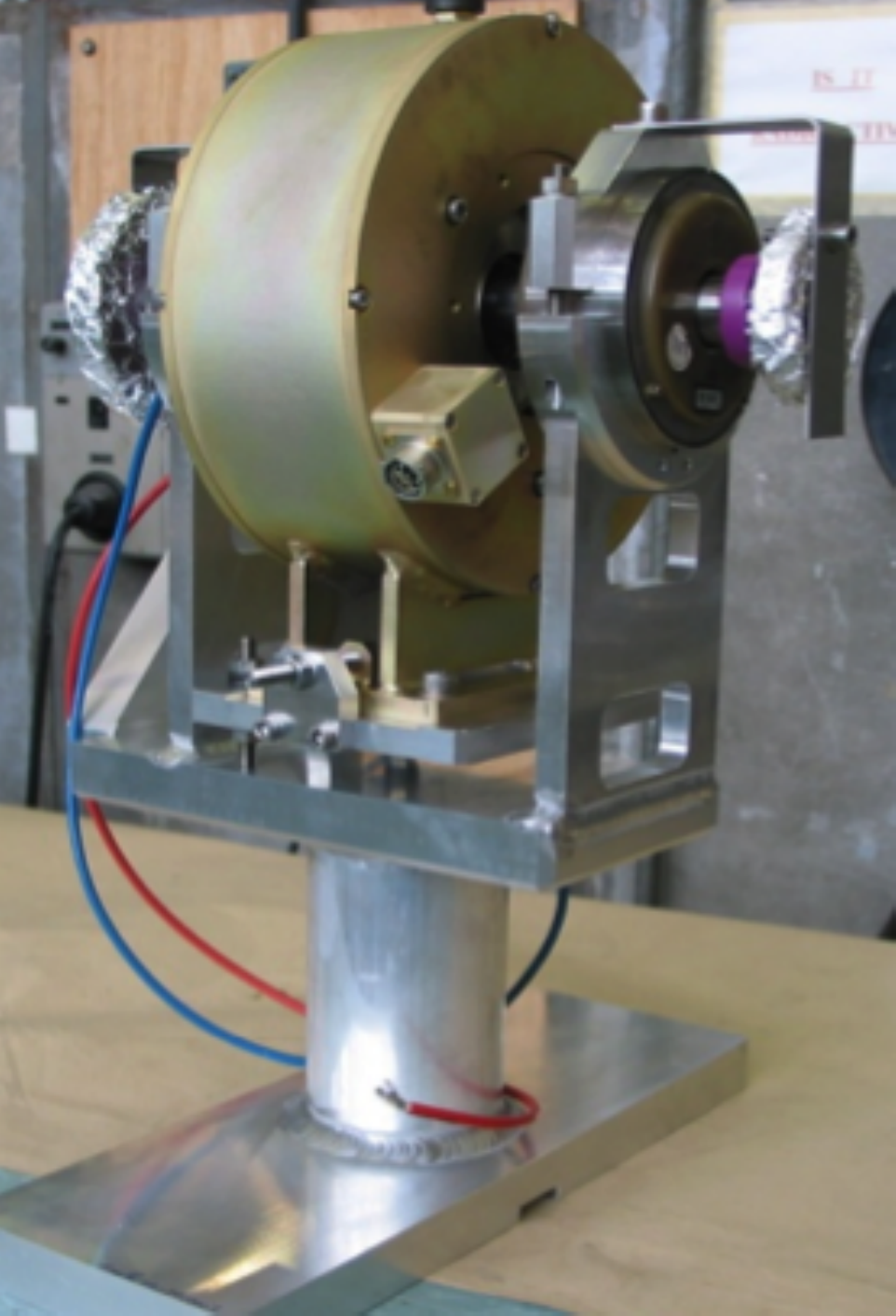}
    \caption{A Beam Current Transformed that can be adapted to measure the muon flux in nuSTORM and, in general, in a Neutrino Factory. Reproduced under CC-BY-SA license with permission from CERN.}
    \label{fig:BCT_nustorm}
\end{figure}

Non-conventional beams like the above-mentioned Neutrino Factories offer interesting opportunities to perform precise flux measurements.
Since in a Neutrino Factory the neutrinos are created by muons stored in a ring, most of the diagnostics of conventional accelerators  can be embedded in such a ring. The nuSTORM Collaboration has envisaged this possibility in Ref.~\cite{Adey:2013pio}. The decay ring is a set of FODO cells (focusing and defocusing quadrupole lattice) and steering magnets that confine the muons ($E_\mu \simeq 5$~GeV) in a circular ring with two long straight sessions (Fig.~\ref{fig:schematics_nustorm}). The circulating muon intensities is measured by a toroid-based Fast Beam
Current Transformer (FBCT). One of these devices has been developed at CERN~\cite{soby} and is shown
in Fig.~\ref{fig:BCT_nustorm}. A one-turn calibration winding
and a 20-turn secondary winding picks up the induced current by the charged particles revolving inside the ring. The windings are wound on a magnetic core and housed in a 4-layer shielding box. This system achieves a precision of $10$~\textmu A in a $10$~MHz bandwidth. It is therefore useful for fast extractions where the muons are bunched in tens of \textmu s up to ms. This class of devices can be used also in other monitored neutrino beams except when the extraction is very long and the induced current goes below the sensitivity of the instrument. In~\cite{Adey:2013pio}, the expected precision on the muon flux is 1\%, which propagate to a precision in $\phi(\nu_\mu)$ and $\phi(\nu_e)$ of similar size since the muon-to-neutrino flux is driven only by the three-body kinematics of the muon decay and the betatron oscillations along the ring. 

\section{Hadron dump diagnostics}
\label{sec:hadron_dump}

Particles in the decay region need to be stopped into a dedicated structure  (hadron dump - Fig.~\ref{fig:narrow_wide}). The hadron dump absorbs the beam stopping all non-decayed hadrons, or particles other than neutrinos, within the thermo-mechanical and radio-protection constraints. It should be noted that in some wide-band beams the hadron dump is also a ``proton dump'' since it is hit by a significant fraction of high-energy primary protons that cross the target without interacting. In this case, the dump requirements are much tighter in terms of particle fluence and irradiation than narrow-band beams.
In narrow-band beams, a large fraction of muons come from the beam halo component having the right momenta to be channeled into the transport line of Sec.~\ref{sec:hadron_beamlines}. This background is a challenge for diagnostics unless it can be separated by exploiting the peculiar transverse and momentum distributions. In general, the production of $\delta$ rays from muons has also to be taken into account to estimate the correct muon flux. 

Muon monitors placed downstream of the hadron dump have been routinely used in many neutrino beams to provide a constraint on the \numu flux and the beam alignment. 
As noted above, they are sensitive to the amount of $\nu_\mu$ produced by $\pi^+ \rightarrow \mu^+ \nu_\mu$ because they monitor the muons collinear to the pions. Hence, these devices may constrain  the leading contribution of $\phi_{\nu_\mu}(E)$.
They provide statistically compelling information even with a single proton pulse and deliver continuous monitoring of the beam conditions. The diagnostics, however, must stand huge rates due to punch-through of interactions in the hadron dump and, in particular, the penetrating component made of neutrons and muons. 

\begin{figure}
    \centering
    \includegraphics[height=6cm]{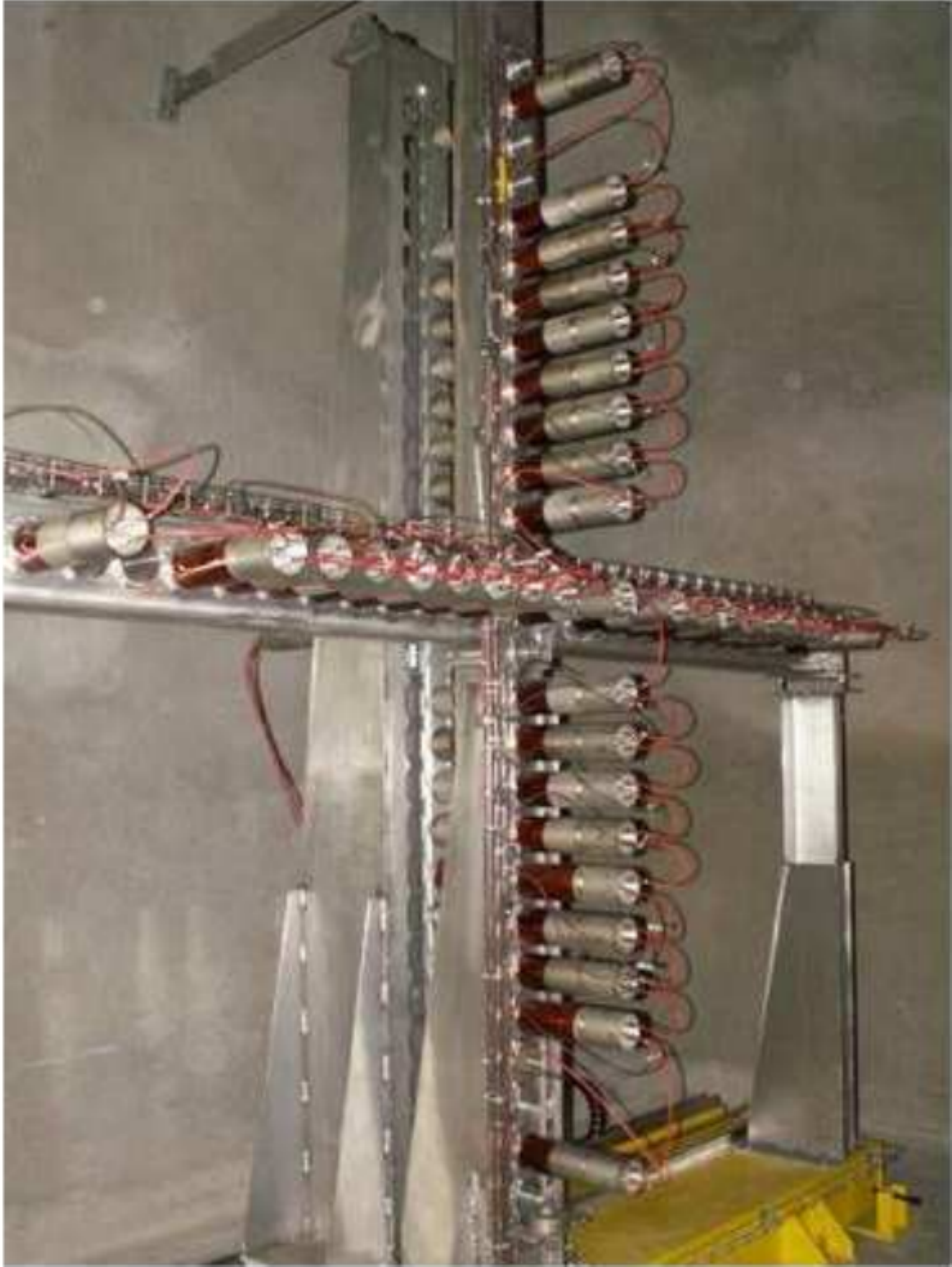}%
    ~~~~
    \includegraphics[height=6cm]{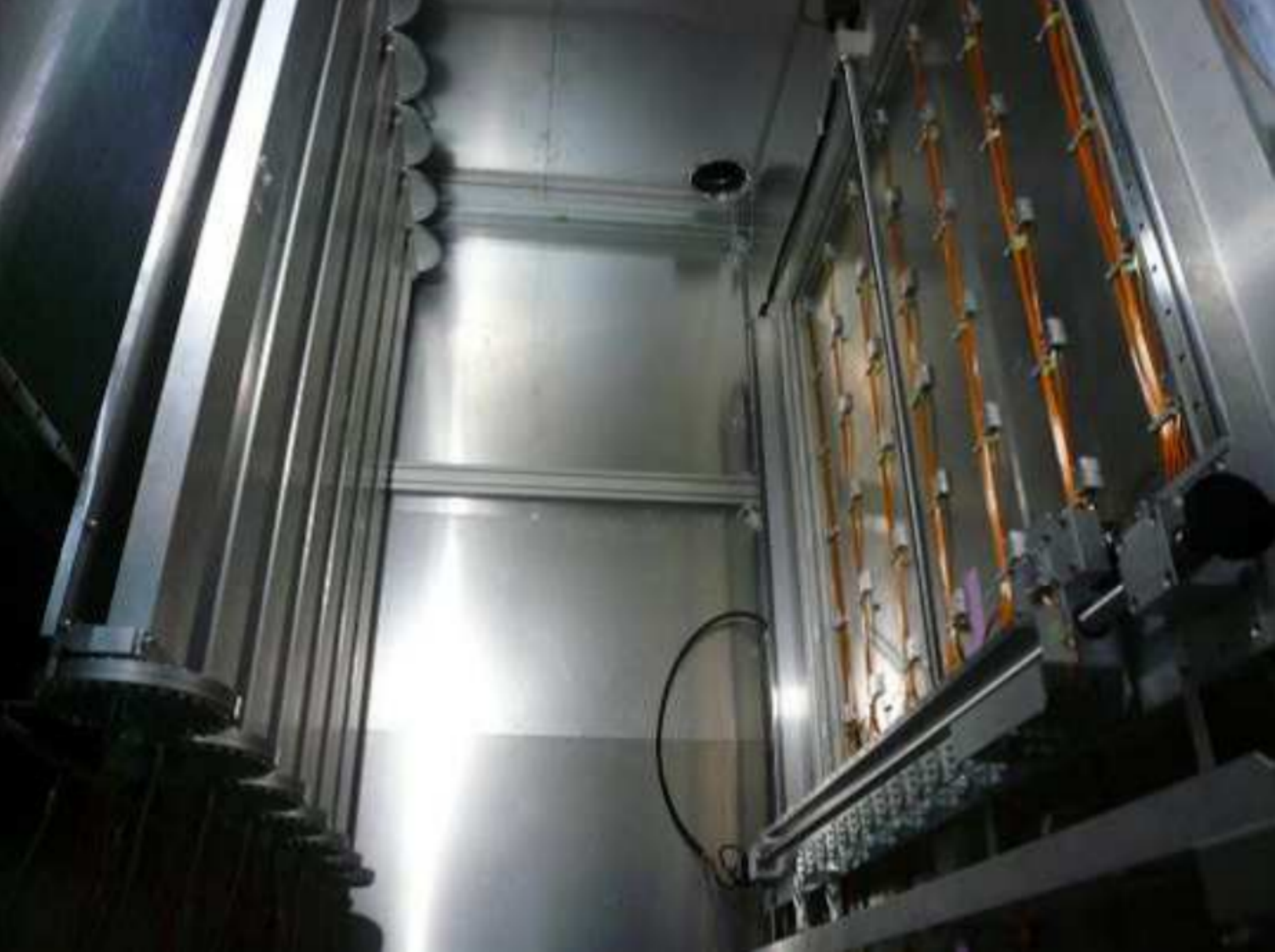}
    \caption{Muon monitoring detectors employed for the CNGS neutrino beam (left) and the J-PARC beam (right). The CNGS chambers consisted of 42 muon detectors separated by $11.25$~cm and spanning a length of $2.7$~m. The J-PARC system (MUMON) is composed of $7 \times 7$ ionization chambers (left side of the picture) and $7 \times 7$, 1~cm$^2$, PIN photodiodes (right side). The beam enters from the right side. Reproduced under CCA4.0-I license with permission from Sissa-Medialab~\cite{CNGSandOPERA}. 
    }
    \label{fig:muchambs}
\end{figure}

 We will briefly recall the solutions implemented at the CERN-PS experiments, IHEP-Serpukhov, CNGS, FNAL-NuMI, and J-PARC. In most cases, the choice of the detector technology has fallen on gas ionization chambers, where the current generated by the large flux of particles can be readout without damaging the detector. More recently, in the J-PARC beam, silicon PIN diodes are being used and the beam designers are considering Electron Multiplier Tubes (EMT~\cite{refEMT}), as well. At the CERN-PS, eight monitoring stations instrumented a $20$~m concrete shielding positioned at the end of the decay volume. At the IHEP-Serpukhov neutrino beam, the iron shielding was sampled using 16 stations, again exploiting ionization chambers. In both cases, the decay volume was short enough to allow a full sampling of the transverse development of the muon beam. Ionization chambers were also used for the muon monitoring of the CNGS beam~\cite{CNGSandOPERA} where two stations composed of a cross-shaped array of $N_2$ filled ionization chambers (Fig.~\ref{fig:muchambs}, left), were deployed downstream of the hadron dump with $67$~m of rock in between ($20$ and $50$~GeV/c threshold for muons).

At the J-PARC beam, the muon monitoring system (MUMON)~\cite{refT2Kmumon} is composed of a hybrid detector comprising $7 \times 7$ ionization chambers and $7 \times 7$, 1~cm$^2$, PIN photodiodes (Fig.~\ref{fig:muchambs}, right).

The muon monitoring system of the NuMI beam~\cite{refNUMImumon} is composed of a hadron monitor, installed at the end of the decay pipe to measure the rates and profile of the surviving beam component and by three muon monitors arrays with a transverse extension of $2.2$~m, placed at about $6$, $18$ and $36$~m downstream of the hadron monitor. 
With this arrangement, the minimal threshold for muons is $5$, $12$ and $24$~GeV for the three stations respectively. Each station consists of arrays of $7 \times 7$ or $9 \times 9$ He-based ionization chambers 
with a gap of 1 or 3~mm for the hadron and muon stations respectively.
Particle rates range, at NuMI, from about $7 \times 10^8$/cm$^2$ for $10^{13}$ protons-per-pulse at the hadron monitor to a few $10^6$/cm$^2$  at the muon monitors. At the higher energies of CNGS, about $10^8$ muons/cm$^2$ are observed during the $10.5$~\textmu s fast extraction.

In narrow-band beams, the possibility to monitor muons fluxes and their spectra downstream of the hadron dump is particularly interesting to access to the bulk of the low-energy muon neutrinos produced from two-body decays of pions. In ENUBET, those are located in the $E_\nu<4$~GeV region of Fig.~\ref{fig:dichromatic}. Unlike kaons, in $\pi^+\to\mu^+\nu_\mu$ the small pion-muon mass difference results in forward-emitted muons that reach the hadron dump before crossing the wall of the decay volume. This flux component is particularly interesting since it lies at lower energies with respect to kaon-generated \numu. 
The ENUBET Collaboration is studying a muon monitoring system composed of 8 stations with iron absorbers ranging from 2~m (upstream) to $0.5$~m (downstream). The transverse distribution of muons is limited to about $30$~cm FWHM and the highest muon fluence is about $2\times 10^6$ muons/cm$^2$.

\begin{figure}
    \centering
    \includegraphics[width=8cm]{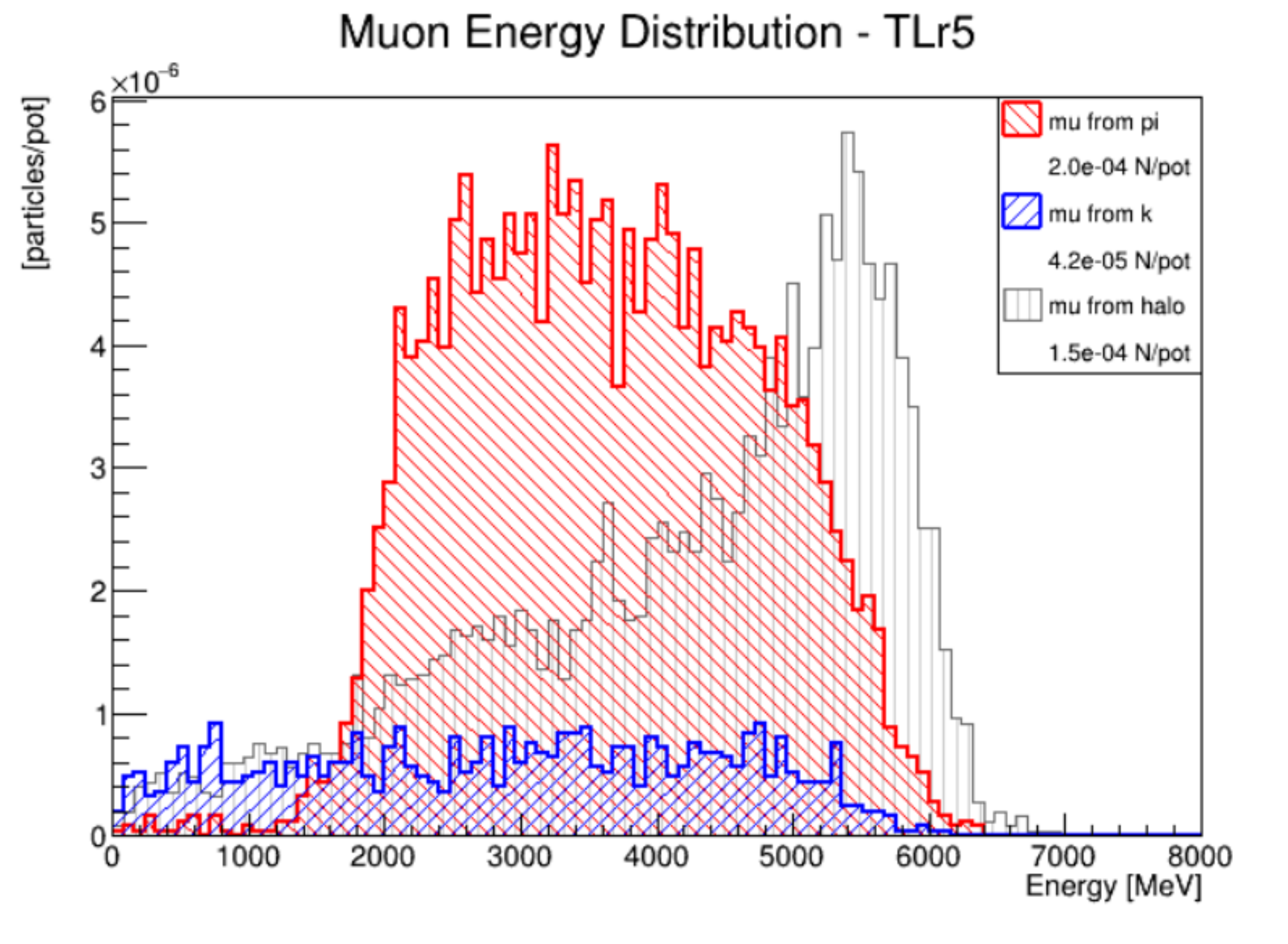}
\caption{Spectra of muons at the ENUBET muon monitors.}
    \label{fig:muchamENUBETsep}
\end{figure}

The total neutron rates integrated over the experiment lifetime at the most upstream station are 
$\sim 8 \times 10^{11}$ decreasing to $6 \times 10^6$ n-1~MeV-eq/cm$^2$
in the most downstream station\footnote{The neutron flux is expressed in non-ionizing-dose units, i.e.~in 1~MeV equivalent neutron fluence (n-1 MeV-eq) and provides the radiation-hardness requirements of the detectors.}. In monitored neutrino beams and all narrow-band beams, radiation damage is much smaller than  wide-band beams, where the hadron dump collects the non-interacting primary protons, too. The designers of monitored neutrino beams envisage the use of fast detectors to reconstruct muons on an event-by-event basis and measure their momentum individually from their range with the detector stations.
The energy and position measurement at the single particle level is a novel tool to separate signal muons from pions and halo muons. The distributions of the two components are well separated in momentum (Fig.~\ref{fig:muchamENUBETsep}) and, to a lesser extent, in the position of the impact point and the reconstructed angle.
The energies of muon neutrinos corresponding to bins of muon momentum - reconstructed by range - show a clear anti-correlation in energy due to the kinematics of 2-body decays. The shape and normalization of the corresponding \numu  are thus precisely constrained by the muon distributions.

Muon diagnostics provide the $\numu$ flux from pions
with a precision of $\sim 20$\% in wide band beams~\cite{numi:flux}. In monitored neutrino beams~\cite{longhin_neutrino2020,terranova_snowmass2020}, it reaches a per-cent level precision and allows for the implementation of the NBOA technique described in Sec.~\ref{sec:nboa}.

\section{Mitigation techniques of beam systematics at the neutrino detectors}
\label{sec:neutrino_detector}

The neutrino detectors of an accelerator beam facility play a dual role: perform the measurements the facility has been designed for (oscillations, cross sections, tests of fundamental symmetries, etc.) and, at the same time, provide information on the energy, flux, and flavor of the neutrinos at the source. The two pieces of information are highly entangled. Modern diagnostics is aimed at decoupling the observables at the neutrino detector, which should be used to perform the physics measurements, and the input from beam monitoring devices, which determine the properties of the source.

\subsection{The Near-Far detector technique}
As mentioned in Sec.~\ref{sec:introduction}, the most powerful decoupling mechanism conceived for the study of oscillations is the construction of two detectors: a smaller detector located in the proximity of the source (Near Detector, ND) and a larger detector located at an optimal baseline to study neutrino oscillations (Far Detector, FD). In its classical implementation, the near detector should be identical to the far detector but located in a position where oscillation effects are negligible. In this way, the Near Detector
provides a measurement of $\nu_\mu$ at $t \simeq 0$ and an estimate of the $\nu_e$ contamination of the beam. The use of identical detectors cancels out systematic differences in the detection efficiency between the ND and the FD and is the technique of choice for moderate-precision disappearance experiments. For instance, if we consider the integral spectrum of $\nu_\mu$ detected at the ND~\cite{Huber:2007em}, the total number of neutrino interactions will be:
\begin{equation}
    N_{ND}= \tilde{M} \phi^{ND}_{\nu_\mu} \epsilon_{\nu_\mu} \sigma_{\nu_\mu}  
\end{equation}
where $\tilde{M}$ is the number of scattering centers, which is proportional to the mass and material of the detector, $\phi^{ND}_{\nu_\mu}$ is the total flux integrated during the data taking at the ND location, and $\epsilon_{\nu_\mu} \sigma_{\nu_\mu}$ is the visible cross section, i.e.~the cross-section corrected by the finite efficiency of the detector. For two identical detectors located sufficiently far from the source\footnote{In this case, the width of the detector is much smaller than the beam width and the flux decreases as the square of the distance.}:
\begin{equation}
    \frac{\phi^{ND}_{\nu_\mu}}{\phi^{FD}_{\nu_\mu}} = \frac{L_{ND}^2}{L_{FD}^2}   
\label{eq:flux_scaling}
\end{equation}
where $L$ is the source-to-detector distance.
In such an ideal condition, the oscillation probability can be inferred straightforwardly by the far-over-near event ratio because:
\begin{equation}
    N_{FD}= \tilde{M}' \phi^{ND}_{\nu_\mu} \frac{L_{FD}^2}{L_{ND}^2} \epsilon_{\nu_\mu} \sigma_{\nu_\mu}  P(\nu_\mu \rightarrow \nu_\mu)  
\end{equation}
and 
\begin{equation}
    \frac{N_{FD}}{N_{ND}}= \frac{\tilde{M}'}{\tilde{M}} \frac{L_{ND}^2}{L_{FD}^2} P(\nu_\mu \rightarrow \nu_\mu)  
\end{equation}
where $\tilde{M}'$ are the scattering centers of the FD. $\tilde{M}'$ is generally $\gg \tilde{M}$ but $L_{FD} \gg L_{ND}$ and the number of events at the ND is much larger than the FD in long-baseline experiments.
Using this simple technique, we can measure $P(\nu_\mu \rightarrow \nu_\mu)$ with $<10$\% precision without relying on sophisticated beam diagnostics. In addition, if the energy of the neutrino can be reconstructed by the final state particles, like in quasi-elastic events where the energy is uniquely determined by the lepton momentum and the scattering angle, the near/far cancellation works for each energy bin and provides the oscillation probability as a function of the energy. This method was mastered by MINOS~\cite{Michael:2006rx} at the time of the discovery of neutrino oscillations and provided some evidence of the oscillation pattern and a solid measurement of the oscillation parameters $\theta_{23}$ and $\Delta m^2_{23}$.  

The near/far cancellation technique dominated the physics of neutrino oscillation in the 90s but has been challenged over the years, when the need for high-precision measurements has emerged. In particular, after the discovery of $\theta_{13}$, this technique turned out to be too coarse for the search of CP violation due to the tiny size of the effects to be observed and the need to control with per-cent precision the entire spectrum where oscillations take place. We will describe this paradigm shift in the sections below.

\subsection{Appearance measurements}

If the facility is aimed at measuring $\nu_\mu \rightarrow \nu_e$ probabilities or its CP conjugate $\nubarmunubare$, substantial changes are necessary to the near/far technique due to the presence of multiple cross sections and efficiencies. 
In this case, we must either improve or cancel out the errors on the $\nu_e$ cross section, the flux and the intrinsic $\nue$ contamination. The simplest way is to employ the near detector to measure the beam contamination. This includes both the intrinsic contamination of $\nu_e$ produced at source (typically a few \%) and the mis-identification background. If the near detector were {\em identical} to the far detector, it would be optimized for \nue detection providing:
\begin{equation}
    N^{e}_{ND} = \tilde{M} \int dE \left[ \epsilon_e(E) \ \phi(\nu_e) \ \sigma_{\nu_e}(E) \ + \epsilon_{\mu \rightarrow e} \phi(\nu_\mu) \sigma_{\nu_\mu}(E) \right]
    \label{eq:near_appearance}
\end{equation}
where $\epsilon_{\mu \rightarrow e}$ is the mis-identification probability of tagging a \numu as a $\nu_e$, e.g.~due to neutral currents (NC) events.
The FD would provide, in this case,
\begin{equation}
    N^{e}_{FD} = \tilde{M}' \int dE \left[ \epsilon_e(E) \ \phi(\nu_\mu) \ \sigma_{\nu_e}(E) P(\nu_\mu \rightarrow \nu_e) \ + \epsilon_{\mu \rightarrow e} \phi(\nu_\mu) \sigma_{\nu_\mu}(E) \right]
    \label{eq:far_appearance}
\end{equation}
and the $N^{e}_{FD}/N^{e}_{ND}$ ratio does not provide any more direct access to the oscillation probability. Note in particular that the ``identical'' near detector has no more access to the product $\phi_{\mu} \epsilon_e \sigma_e$, which is essential to estimate the signal at the far detector:
\begin{equation}
    \phi(\nu_\mu) \epsilon_{e} \sigma_{\nu_e} P(\nu_\mu \rightarrow \nu_e)
\end{equation}
The simplest fix is to have identical detectors that are able to measure both \numu and $\nu_e$, gaining two additional observables:

\begin{equation}
    N^{\mu}_{ND} = \tilde{M} \int dE \left[ \epsilon_\mu(E) \ \phi(\nu_\mu) \ \sigma_{\nu_\mu}(E) \ + \epsilon_{e \rightarrow \mu} \phi(\nu_e) \sigma_{\nu_e}(E) \right] \simeq \tilde{M} \int dE \left[ \epsilon_\mu(E) \ \phi(\nu_\mu) \ \sigma_{\nu_\mu}(E) \right]
    \label{eq:near_appearance_mu}
\end{equation}
The last approximation works because $\phi(\nu_\mu) \gg \phi(\nu_e)$ in any conventional beam. The second observable is:
\begin{gather}
    N^{\mu}_{FD} = \tilde{M}' \int dE \left[ \epsilon_\mu(E) \ \phi(\nu_\mu) \ \sigma_{\nu_\mu}(E) P(\nu_\mu \rightarrow \nu_\mu) \ + \epsilon_{e \rightarrow \mu} \phi(\nu_e) \sigma_{\nu_e}(E) P(\nu_e \rightarrow \nu_\mu) \right] \simeq  \nonumber \\ \tilde{M}' \int dE \left[ \epsilon_\mu(E) \ \phi(\nu_\mu) \ \sigma_{\nu_\mu}(E) P(\nu_\mu \rightarrow \nu_\mu) \right]
    \label{eq:far_appearance_mu}
\end{gather}
In particular, if the mis-identification probability $\epsilon_{\mu\rightarrow e}$ can be established independently using beam test, cosmic ray runs, control samples, etc., or the background can just be neglected, then  $N^{ND}_e$ provides a pure measurement of the beam contamination as seen by the far detector. $N^{ND}_\mu$ measures $\epsilon_\mu(E) \ \phi(\nu_\mu) \ \sigma_{\nu_\mu}(E)$ as seen by the far detector and the main systematics are now on the ratio:
\begin{equation}
 \frac{\epsilon_e \sigma_{\nu_e}}{ \epsilon_\mu \sigma_{\nu_\mu} }  
\end{equation}
This is the reason why the current generation of oscillation experiments use the near/far technique mostly to evade the large uncertainties on the flux, relying on the lepton universality ($\sigma_{\nue} \simeq \sigma_{\numu}$) and the measurements with cosmics or particle beams to mitigate the uncertainties on $\epsilon_e/\epsilon_\mu$. This technique has been perfected by T2K and \nova reaching a systematic budget of 8-5\%. 

Clearly, the near/far technique cannot be used for cross-section experiments, where the detector is always located at a short distance and the (poorly known) flux just scales as $L^2$ at large distances. 

\subsection{High precision NDs}
\label{sec:ND}

Even if the near/far cancellation worked perfectly for a monochromatic neutrino beam, it would fail for a real beam where the spectrum extends in a broad band. The main reason is that the ND records the entire flux because its dimension is comparable with the dimension of the beam at $L_{ND}$, while the FD traces only the forward part of the beam since it covers a tiny solid angle. At the location of the FD, the transverse size of the neutrino beam is much larger than the size of the detector. The actual region seen by the FD thus depends on both $L_{ND}$ and the off-axis angle of the FD. The net result is that the spectrum at ND is different from the spectrum at FD and we cannot rely anymore on Eq.~\ref{eq:flux_scaling}. 

This challenge is being faced in two ways by modern long-baseline experiments. In the last ten years, the ND complex has significantly increased in complexity. A quite radical view was pursued by T2K building the ND280 and INGRID detectors~\cite{Abe:2019vii,Abe:2019whr}. INGRID is a simple scintillator-iron detector that monitors the relative changes of the beam on-axis and provides an overall normalization.
The ND280 is located off-axis along the path to the FD, i.e.~Super-Kamiokande~\cite{Fukuda:2002uc} (SuperK). It is very different from SuperK because the neutrino rate is too high to be properly reconstructed with a water Cherenkov detector but, on the other hand, has a much better granularity and PID capability. More recently, the detector is being upgraded to increase the angular acceptance up to $>90^\circ$. In this way, ND280 can reconstruct in detail the beam composition and spectrum to reproduce the phase space visible at the FD with a precision that would be inconceivable for Super-K. It identifies the potential background directly measuring the particles that escape detection in water. Notable examples are NC events where the pions are below the threshold for Cherenkov light production, NC+$\pi^0$ events and events where the final state hadrons re-scatter or are re-absorbed inside the nucleus. All this information corrects the neutrino interaction simulation in a data-driven model, which is then employed to estimate the actual efficiency and purity of the SuperK \nue charged-current (CC) sample. 

The price to pay is a substantial increase of complexity because the Monte Carlo systematics are cured by ND beam data and we give up the ``two-identical detector'' principle.  Ten years ago there was some skepticism on the outcome of this approach and ND280 was supposed to be integrated by an {\em intermediate detector} identical to SuperK and a high-precision tracker based on Liquid Argon. These doubts were superseded during the run of ND280, which achieved a flux systematic $< 8$\%. The task has been further eased by the unexpectedly large size of $\theta_{13}$ that makes the signal rate ten times larger than the estimate of the T2K Proposal.

\subsection{Energy unfolding}

A high precision ND is also an excellent tool to correct for the spectral shape of the beam. As noted above, the beam is not monochromatic and the spectrum at FD is quite different than the spectrum at ND. To correct for the mismatch, the ND must rely on a good reconstruction of the neutrino energy. Unlike the quasi-elastic events mentioned above, the reconstruction of the energy cannot be performed analytically and depends on the actual number of final state particles that are reconstructed.
Particle losses or mis-identification contribute to biases that affect both the long-baseline oscillation experiments~\cite{Benhar:2015wva} and the short-baseline cross-section experiments~\cite{katori2018,Formaggio:2013kya,atar2020}. The main source of uncertainty comes from final-state interactions of particles inside the nucleus, undetected neutral particles, and particles produced outside the acceptance of the detector. Two novel techniques have been recently proposed to overcome this limitation: the PRISM concept for long-baseline experiments and the narrow-band off-axis technique for short-baseline experiments. Both techniques exploit some a priori knowledge of the neutrino energy coming from the source to sidestep the use of final state particles and reduce the biases.

\subsection{The PRISM technique}

The PRISM concept~\cite{Bhadra:2014oma} is based on a movable ND located in the proximity of the source. The motion can be in any axis perpendicular to the beam axis. Both DUNE and HK envisage the use of this method: DUNE with an ND complex moving in the horizontal plane, HK with a water Cherenkov detector moving vertically in a shaft. In this way, the ND can record the spectrum as a function of the off-axis angle $\theta$. Due to the two-body kinematics, a neutrino emitted at an angle $\theta$ with respect to the beam axis and produced by a pion decaying along the beam axis has an energy of:
\begin{equation}
    E_{\numu} = \frac{m_\pi^2-m_\mu^2}{2(E_\pi-p_\pi \cos \theta)}
\label{eq:off-axis_concept}
\end{equation}
As a consequence, for any $\theta$ we can find the maximum neutrino energy differentiating Eq.~\ref{eq:off-axis_concept}. This energy is:
\begin{equation}
    E_{\numu}^{Max} = \frac{m_\pi^2-m_\mu^2}{2 E_\pi \sin \theta}
\label{eq:off-axis_concept_2}
\end{equation}
The pion distribution is not monochromatic, and the wider the band of the beam, the wider the number of neutrinos at an angle $\theta$ that have an energy far from $E_{\numu}^{Max}$.

PRISM traces the energy spectrum of the wide-band beam as a function of $\theta$. Each detector position provides a spectrum and the beam is sliced in many distributions, whose mean energy is given by Eq.~\ref{eq:off-axis_concept_2} with a width smaller than the original band. In a sense, a single PRISM point records an approximate monochromatic beam where the energy of the incoming neutrino is known a priori with a smaller width. All these slices are thus combined to provide the correct
migration matrix at FD, i.e.~the probability that a \numu with energy $E$ is seen at a reconstructed energy of $E'$ due to the biases of the final state particles. This approach mitigates the problem of energy unfolding and the corresponding systematics on the oscillation parameters.

\subsection{The narrow-band off-axis (NBOA) technique}
\label{sec:nboa}

\begin{figure}[h]
\centering
\includegraphics[scale=0.7]{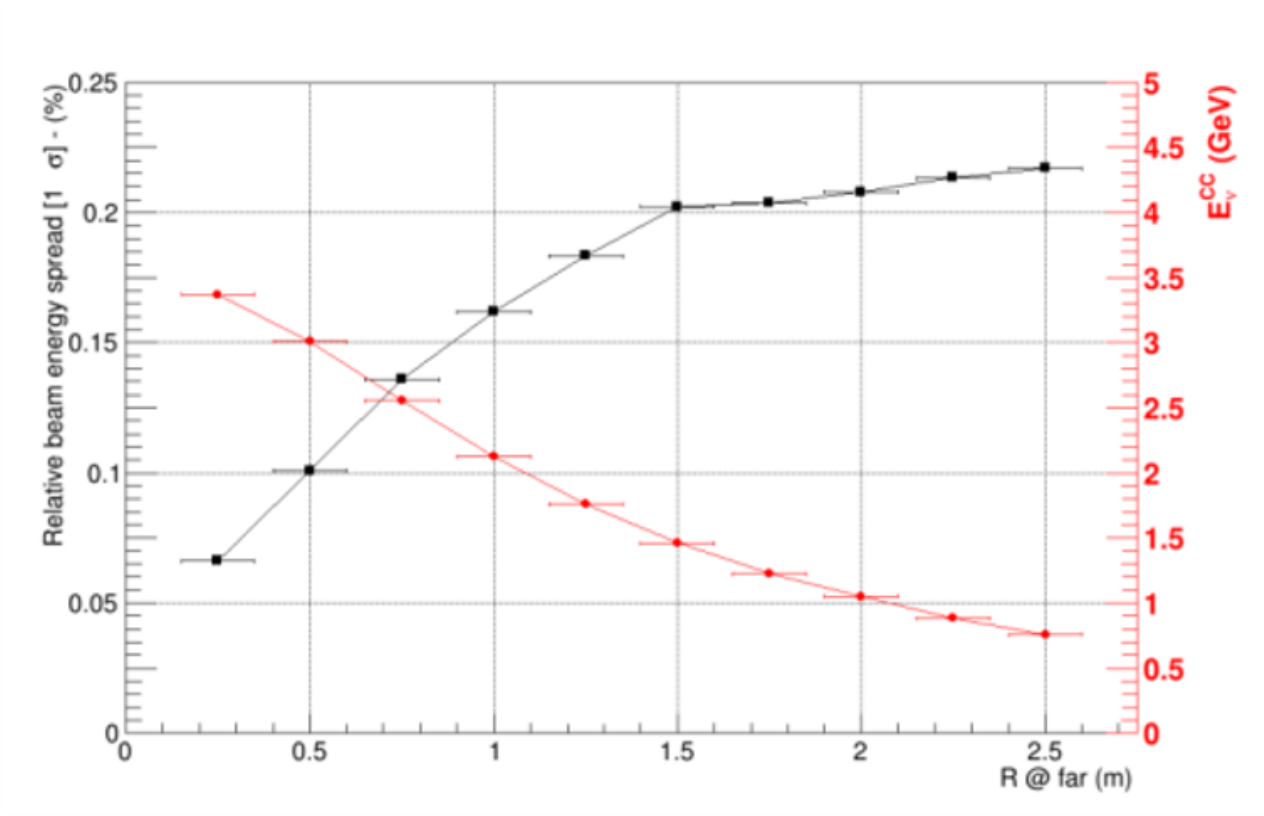}
    \caption{Black points:  neutrino energy resolution, corresponding to the $1 \sigma$ beam width, versus the distance of the $\nu_\mu$ interaction vertex to the beam axis (``R @ far''). Red points: mean neutrino energy (in GeV) versus R @ far. }
    \label{fig:nboa}
\end{figure}

A technique similar to PRISM has been proposed in~\cite{ENUBET_proposal} for short-baseline experiments and, in particular for the cross-section measurements and the study of oscillation patterns in sterile neutrinos. The PRISM technique is mostly conceived for high-intensity long-baseline experiments and requires movable detectors of moderate mass (500 tons or more). Its implementation is too costly and complex for a cross-section experiment, whose anticipated investment is two orders of magnitude smaller than DUNE or HK. Since the ideal setup for a monitored neutrino beam is a narrow-band beam, the intrinsic width around $E^{Max}$ (see Eq.~\ref{eq:off-axis_concept_2}) is much smaller than a wide band. This width creates a much tighter correlation between the neutrino emission angle $\theta$ and its energy.
In particular, the ENUBET collaboration has recently shown that the measurement of $\theta$, which corresponds just to the identification of the interaction vertex of the detector, provides the energy with a precision that improves with the energy itself. For the current ENUBET design (mean secondary momentum at $8.5$~GeV/c and $10$\% momentum bite), the precision ranges from $8$\% at $3$~GeV to $20$\% at $1$~GeV (see Fig.~\ref{fig:nboa}). This is definitely a valuable result: at high energy (e.g.~for the DUNE cross-sections), $E_\nu$ is known very precisely and practically all biases due to energy reconstruction are removed. At low energy, the events are low-multiplicity elastic scatterings or events with resonant/coherent production of mesons. In this case, the missing particles in the energy reconstruction produce an inconsistency between the a priori measurement of $E_\nu$ from the NBOA and the measurement based on final state particles. For pure quasi-elastic events, where this inconsistency never occurs, the precision is dominated by the detector measurement through the usual quasi-elastic (QE) formula~\cite{Nieves:2012yz}:
\begin{equation}
    E_{rec} = \frac{M E_\mu -m_\mu^2 /2}{M-E_\mu + |\mathbf{p}_\mu| \cos \theta_\mu} 
\end{equation}
where the nucleon of mass $M$ is at rest in the laboratory frame (Fermi motion and nuclear effects are neglected) and $(E_\mu, \mathbf{p}_\mu)$ is the four-momentum of the outgoing muon. 

\subsection{Flux measurement at the detector}

As mentioned in Sec.\ref{sec:ND}, the ND cannot measure the flux in a direct manner because it cannot disentangle the product $\phi(E) \epsilon(E) \sigma(E)$ neither for $\nu_e$ nor for $\nu_\mu$. A possible way out that can be implemented in high-intensity wide-band beam is the use of the neutrino elastic scattering on the electrons of the detector medium~\cite{Valencia:2019mkf,Marshall:2019vdy}:
\begin{equation}
    \nu_\mu + e^- \rightarrow \nu_\mu + e^-
\label{eq:qe-ele}
\end{equation}
The corresponding cross section is purely leptonic and free of QCD/nuclear effects. It is one of the cleanest observables of the Standard Model because is produced by
the exchange of a $Z$ boson and only depends on the couplings between the $Z$ and the electron. It is, therefore, known with outstanding precision and can be used as a ``standard candle'' to estimate the integrated flux and, with poorer precision, the spectral flux $\phi_{\numu}$. 

\begin{figure}[H]
\centering
\includegraphics[width=8 cm]{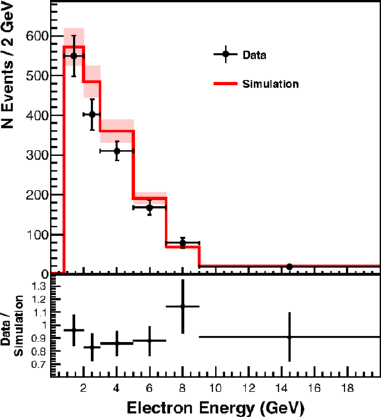}
\caption{Reconstructed electron energy in $\numu e^- \rightarrow \numu e^-$ after background subtraction
and efficiency correction in data and simulation (above)
and the ratio of data to simulation (below). The error bars on the data
include both statistical and systematic uncertainties. Reproduced under CCA4.0-I licence with permission from the American Physical Society~\cite{Valencia:2019mkf}.}
\label{fig:minerva_escattering}
\end{figure}

The signature of Eq.~\ref{eq:qe-ele} is an isolated electromagnetic shower appearing in the same direction as the neutrino beam. For MINER$\nu$A~\cite{Valencia:2019mkf}, no background mips must be found in a $10^\circ$ cone around the candidate electron. The dominant background in these analyses is always the \nue and \nubare CC events with a missing neutron. This background is mitigated by cuts in the $(E_e,\theta_e)$ plane, where $E_e$ is the electron energy and $\theta_e$ is the electron angle with respect to the beam axis. The selected sample in MINER$\nu$A is shown in Fig.~\ref{fig:minerva_escattering}.

\noindent
Using this method, MINER$\nu$A reduced the integrated flux uncertainty from $7.6$\% to $3.9$\% and similar performance are expected in DUNE~\cite{Marshall:2019vdy}.

The electron-scattering technique has three important drawbacks. Firstly, the cross section is extremely small: $10^{-3}$ smaller than the nucleon scattering cross section. This implies that the measurement does not work for low-intensity beams but is an asset for DUNE and HyperK. For cross section measurements, it can be used only as an ancillary measurement to be combined with the flux constraints of monitored neutrino beams. Secondly, the topology is just an isolated electron around the axis of the beam: background soon becomes intractable if the pile-up rate is high or the $\nu_\mu$ background is not known with very high precision. In turn, this background introduces additional systematics that come from nuclear effects in neutrino-nucleon scattering. Finally, it is worth mentioning that conventional beams are completely dominated by the \numu flux and, therefore, this technique does not provide information on the \nue flux, which must be inferred either by hadron production+simulation or by lepton monitoring as in ENUBET.  

Another technique, developed by the CCFR collaboration in the 80s, is based on the selection of \numu CC events ($\nu_\mu +A \rightarrow \mu^- + X$) where the energy of the system $X$ recoiling against the muon is very small. These events are called ``low-$\nu$'' interactions because the kinematic variable $\nu$ is defined as the sum of the energy of all particles belonging to $X$ ($\nu$ is then the hadronic energy $E_X$). The corresponding cross section for deep-inelastic events (DIS) is:
\begin{gather}
    \frac{d\sigma_{\nu_\mu}}{d\nu} = \frac{G_F^2 M}{\pi} \left( \int_0^1 F_2 dx -\frac{\nu}{E_\nu} \int_0^1 \left[ F_2 +x F_3 \right] dx \right. \nonumber \\
    \left . + \frac{\nu}{2E_\nu^2} \int_0^1 \left[ \frac{Mx(1-R_L)}{1+R_L} F_2 \right] dx + \frac{\nu^2}{2E_\nu^2} \int_0^1 \left[ \frac{F_2}{1+R_L} +x F_3 \right] dx \right)
    \label{eq:lownu}
\end{gather}
where $G_F$ is the Fermi constant, $M$ is the proton mass, $x$ is the Bjorken's scaling  variable and $F_{1,2,3}$ are the structure functions. $R_L$ is a function of the previous variables and is proportional to the longitudinal structure function $F_L$~\cite{Bodek:2012uu}. Equation~\ref{eq:lownu}
indicates that for $\nu \rightarrow 0$,  i.e.~for $\nu$ smaller than a given threshold $\nu_{thr}$, the cross section is independent of the neutrino energy $E_\nu$. As a consequence, the measurement of the low-$\nu$ interaction rate $N(\nu<\nu_{thr},E_\nu)$  as a function of the neutrino energy is equivalent to a measurement of the shape of the neutrino flux because:
\begin{equation}
 N(\nu < \nu_{thr},E) \sim \phi(E_\nu) \int_0^{\nu_{thr}} \frac{d\sigma}{d\nu} = \phi(E_\nu) \sigma(\nu < \nu_{thr},E_\nu) \sim \phi(E_\nu)  
\end{equation}
Even if this method only provides the spectrum of the flux and is plagued by the systematics associated with the choice of $\nu_{thr}$, it provides useful information, especially in the energy region where the systematics due to the beamline simulation are particularly high. Besides, 
$\sigma(\nu < \nu_{thr},E_\nu)$ is much larger than the corresponding $\nu_\mu-e^-$ cross-section and statistics is not an issue even in low-power beams. The technique works provided that Eq.~\ref{eq:lownu} holds, i.e.~the experiment records DIS events. This is the reason why it has been used mostly in high $E_\nu$ experiments like NuTev~\cite{Zeller:2001hh}, NOMAD and MINER$\nu$A, and may be particularly useful for DUNE.

\section{From monitored to tagged neutrino beams}
\label{sec:tagged}

The possibility of measuring the time of neutrino interactions at the $\mathcal{O}(100\text{~ps})$ level would open up new opportunities to further improve the precision of neutrino beams.
The {\em stroboscopic beam} is a proposal that goes in this direction~\cite{bunchtiming}.
The arrival time of neutrinos inherits the radio-frequency bunch time-structure of the protons with smearing due to the spectrum
of the energies of secondary mesons: in practice, neutrinos generated from low-energy hadron parents will arrive later at the detector and high-energy parents will produce neutrinos that arrive before. 
We can select samples with markedly different energy spectra  by picking-up neutrino interactions according to their arrival time. The fractions of events of each flavor have some dependence on the arrival time, too, as well as the fraction of events originating from different parents (pions, kaons and muon DIF).
 The achievable performance in an upgraded FNAL Main Injector ring with superconducting radio-frequency cavities capable to re-bunch the present $53.1$~MHz by a factor of 10 is discussed in~\cite{bunchtiming}.

Improving the time tagging accuracy of monitored neutrino beams down to $\mathcal{O}(100\text{~ps})$ in the decay volume would represent a major breakthrough. The reconstruction of leptons on an event-by-event basis gives us the possibility to time-correlate neutrino interactions with the decay products, i.e.~associate every neutrino interaction seen in the neutrino detector to the corresponding charged-lepton seen in the decay volume. This facility was envisaged in the 60s and is called a {\em tagged neutrino beam}~\cite{Hand1969,Pontevcorvo1979}.
A precise time-coincidence resolves the flavor of the neutrino a priori, i.e.~without inferring it from the reconstruction of the lepton produced in CC interaction\footnote{Non-standard neutrino interaction effects could be studied in this way by comparing the lepton emitted at the production and interaction stage.}. If we are able to reconstruct also the other  decay products of the parent ($\pi$ or $K$), we can constrain the energy of the neutrino with even  higher precision than NBOA. The time-tagging would, hence, raise the purity of the selected sample of neutrino interactions to an unprecedented level.

Once a neutrino interaction is observed in the neutrino detector, we can select all lepton candidates whose recorded time is compatible with the neutrino time within the resolution of  the detectors.
Neglecting the neutrino mass, the time of the lepton time-tag ($t_l$) and of the neutrino interaction
($t_\nu$) are linked by:
\begin{equation}
c(t_{\nu} -t_{l}) = d_{di}-d_{dt}\frac{E_l}{p_l}
\label{eqtag}
\end{equation}
The distance between the position of the decay and the lepton interaction in the decay volume ($d_{dt}$)
and the distance between the decay and the neutrino interaction ($d_{di}$) depend on the radial position of the decay and the interaction in the transverse plane and on the emission angles of the lepton and the neutrino. In a two-body decay, where the lepton
and the neutrino are coplanar, we have:
\begin{gather*} 
d_{dt}=(r_{t}-r_{d})/\tan\theta_{t} \\ d_{di}=(r_{i}-r_{t})/\tan\theta_{\nu}
\end{gather*}
where $d$, $i$, $t$ indicate the decay, the neutrino interaction and the lepton tag, respectively. For 3-body
decays, the azimuthal angles also play a role.
Since the neutrino production vertex is unknown, the correction term
given by $d_{di}-d_{dt}\frac{E_l}{p_l}$ in Eq.~\ref{eqtag} can only be approximated by $z_{i}-z_t$ i.e.~the distance of the neutrino
interaction vertex ($z_i$) and the lepton tag ($z_t$) projected along the axis of the decay volume. This approximation causes a systematic time-shift due to the fact that the lepton and the neutrino are not perfectly collinear.
This term introduces an ``irreducible'' time-spread ($\delta_{irr}$), which can exceed $100$~ps in some facilities. This is the reason why $100$~ps is a natural goal for timing in monitored neutrino beams.
This effect can be mitigated by constraining the position of the decay vertex, i.e.~by reconstructing at least two decay products.
The time matching condition reads:
\begin{equation}
    \Delta t = \vert t_\nu -t_{l} -(z_i-z_{l})/c \vert  < \delta_{tagger}\oplus\delta_{\nu-det}\oplus\delta_{irr}.
\end{equation}
The rate of accidental coincidences is the main background of tagged neutrino beams. They arise when a mis-identified lepton is randomly associated because the true lepton is either out of the acceptance of the decay volume instrumentation or lost by inefficiencies in PID. Furthermore, if the neutrino is produced outside the decay volume, a time match can only be due to random coincidences.
If we label with $\delta$ the combined time resolution of both detectors ($\delta_{tagger}\oplus\delta_{\nu-det}$) and the intrinsic smearing ($\delta_{irr}$) due to non-collinearity, the rate of accidental coincidences is proportional
to the product of $\delta$ and the rate of fake lepton candidates that enter the time window. To increase the purity of time matching, the combined time resolution must be smaller than $\delta_{irr}$. 
Compared with monitored neutrino beams, the number of true
time-coincidences can be enhanced by increasing the geometrical acceptance of the detectors in the decay volume and their efficiency, and reduce the number of neutrinos produced outside this volume (upstream decays or $K^0_L$ decays near the target).
Detector technologies for time-tagged neutrino beams must provide excellent timing over large areas. The baseline option of ENUBET is given by the plastic scintillators of the photon veto and has a time resolution of about $400$~ps over large areas. Better options are being considered in the context of a dedicated project (NUTECH~\cite{NUTECH}) exploring different detector technologies such as fast Micromegas with Cherenkov radiators (PICOSEC~\cite{PICOSEC}), Large Areas Picosecond PhotoDetectors (LAPPD\cite{LAPPD}) or LYSO(Ce) crystals.

\section{Conclusions: a step beyond the state-of-the-art}
\label{sec:conclusions}

In the previous chapters, we reviewed the main components of an accelerator neutrino beam, the techniques for diagnostics and the barriers that hinder a per-cent level precision. Such precision is needed for the next generation of long-baseline experiments (DUNE, HK and other proposed facilities) and short-baseline experiments (for cross-section measurements and neutrino physics beyond the Standard Model). 

The reader may wonder what are the leading systematics that jeopardize high-precision diagnostics and the most effective counter-measurements. We can illustrate it considering the world best measurement of the flux performed in the $1$-$10$~GeV range by MINER$\nu$A in 2016~\cite{numi:flux}.

\noindent
This analysis used one of the largest datasets of ancillary measurements both off-site and on-site:
\begin{itemize}
    \item the MIPP data on thick Carbon target~\cite{PhysRevD.90.032001}
    \item the NA49 and NA61/SHINE data on thin Carbon target~\cite{Alt:2006fr,Baatar:2012fua,PhysRevC.84.034604}
    \item the muon chambers located after the beam dump
    \item the neutrino-electron elastic scattering at the MINER$\nu$A detector
    \item the low-$\nu$ neutrino-nucleus scattering at the MINER$\nu$A detector
\end{itemize}
combined with beam diagnostics: the alignment tolerance on the beamline elements~\cite{numi:beam,zarkothesis,Zwaska}, the horn current ($180\pm 2$~kA) and the detector geometrical parameters. The POT were estimated by two toroids located upstream of the target with a $0.5$\% precision~\cite{numi:beam}. The beam losses were monitored by sealed-gas ionization chambers and four total loss monitors\footnote{These devices are coaxial cables filled with a Ar-CO$_2$ mixture. They produce a signal proportional to the ionization caused by protons lost before the target.}. Like ENUBET, the NuMI designers installed three layers of muon chambers after the beam dump to estimate the energy and flux of the $\mu^+$ from $\pi^+$ decay. However, since the decay volume is located on axis with the horn, the environment is much harsher and does not provide muon monitoring at the single-particle level.
Ionization chambers were also installed just before the dump to make a rough estimate of the total undecayed hadrons and, especially, the beam spot size. In this case, the rate is extremely high: up to $10^9$ charged particles/cm$^2$/spill ($10^{14}$~Hz/cm$^2$) during beam
operation, consisting mainly of $120$~GeV/c primary protons
that have not interacted in the beamline. These monitors are crossed by $2\times  10^9$ neutrons/cm$^2$/spill originated by the albedo in the beam dump. The particle fluence at the center of the monitor is also very large: $1.3$~Grad/year. These rates support the need to decouple the proton and hadron dump in high-precision beams like the monitored neutrino beams. 

\begin{table}[H]
\caption{Diagnostics in long-baseline (LB), short-baseline with monitored neutrino beams (SB-MNB) and Neutrino Factories (NF). The leading diagnostic tool driving the systematic budget is labeled ``critical''. The tools that are not needed because sidestepped, for instance by the lepton monitoring, are labeled ``irrelevant''. BCT in TL means a Beam Current Transformer located in the transfer line. ``$\simeq$ irrelevant'' describes cases where the information is not needed at leading order but is useful to estimate second-order effects like kaon (muon) decays before the decay volume (storage ring) in MNB (NF). We labeled ``OK'' the diagnostic tools that are needed but do not limit the current precision on the flux. }
\label{tab:techniques}
\centering
\begin{tabular}{cccc}
\toprule
\textbf{Source}	& \textbf{LB}	&  \textbf{SB-MNB} &\textbf{NF} \\
\midrule
POT		& OK			& irrelevant & irrelevant \\
secondary yield		& \textbf{critical}			& ancillary & ancillary \\
transport		& $\simeq OK$	& $\simeq$irrelevant		& $\simeq$irrelevant \\
muon monitoring		& marginal &	\textbf{critical}		& ancillary \\
$\nu$ detector		& \textbf{critical}			& ancillary & ancillary \\
lepton monitoring (MNB) & not used & \textbf{critical} & ancillary \\
BCT in TL & not used & ancillary & \textbf{critical} \\ 
\bottomrule
\end{tabular}
\end{table}

The entire NuMI beamline is simulated with GEANT4 and the external data from MIPP and NA61/SHINE are introduced as weights in the theoretical distributions of the hadron yields. Tertiary production in the beamline or secondary meson absorption is simulated with GEANT4 correcting the cross section by external data coming mainly from T2K~\cite{t2k:flux}.
Measurements in the MINER$\nu$A neutrino detector are used both as a check of flux predictions and as an additional constraint to the flux.

The results achieved at NuMI show that classical diagnostics (GEANT4/FLUKA simulation with beam monitors, including the  muon chambers) can achieve a precision of $10$-$25$\% depending on the energy range. The hadron production data reduce such uncertainty to $5.4$\% and completely dominate the  current systematic uncertainty of NuMI and most of the other neutrino beams. Muon chambers at such a high rate play a minor role to constrain the flux but monitor relative flux variations during the data taking. The uncertainties on the beamline geometry and the horn current are generally negligible with respect to hadron production except in the falling-edge region of the flux. 

\begin{figure}[h]
\centering
\includegraphics[scale=0.5]{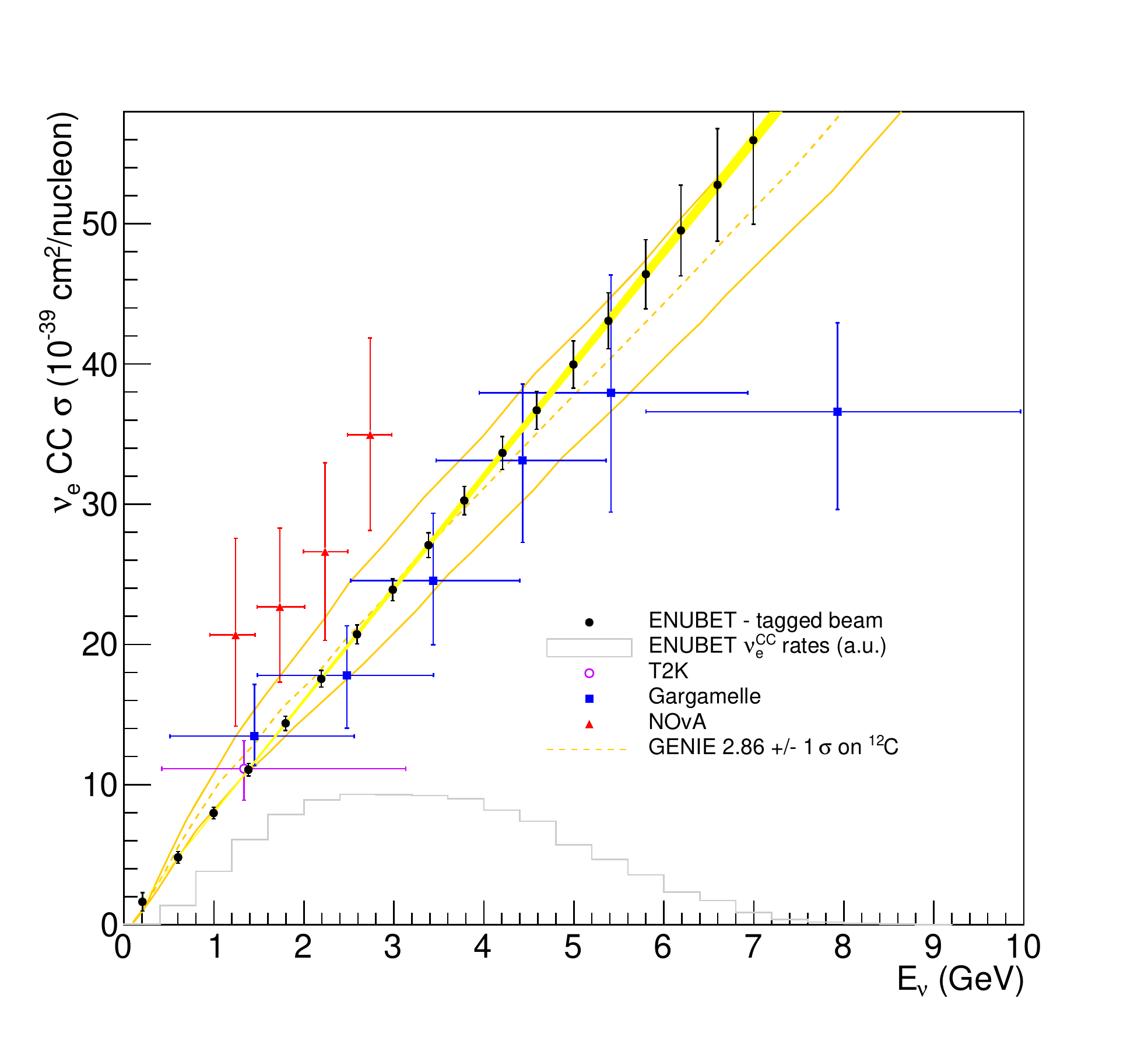}
    \caption{Impact of ENUBET data on the \nue cross-section compared with current measurements from T2K, \nova and Gargamelle. Theory expectations (dashed orange line) and the corresponding systematics (region between the orange lines) are shown together with the overall systematic budget of ENUBET (yellow band). The gray line shows the energy spectrum (in a.u.) of the ENUBET beamline tuned for the DUNE energy range.}
    \label{fig:enubet_xsect}
\end{figure}

All these considerations support the improvements of hadron production experiments for high-intensity long-baseline experiments. For short-baseline facilities and, in particular, in cross-section experiments, the method of choice should be the monitored neutrino beam, which sidesteps all systematics before the decay volume. In practice, it is wise to combine lepton monitoring with hadron production data to cross-check the flux predictions.
Neutrino detector measurements are useful, especially if the beam intensity is very high like in the DUNE and HyperK NDs, to gain additional constraints on the $\numu$ flux down to a 5\% precision level~\cite{bravar_nufact}.
Table~\ref{tab:techniques} summarizes the techniques that can be perfected for the next generation of experiments and the leading systematics. As a final example, Fig.~\ref{fig:enubet_xsect} shows the anticipated contribution of NP06/ENUBET to the measurement of the total electron-neutrino cross section compared with current data assuming a 1\% precision on the flux.

\vspace{6pt} 



\authorcontributions{Conceptualization and writing--original draft preparation, N.C. (Sec.~\ref{sec:anb},\ref{sec:proton},\ref{sec:hadron_beamlines}), A.L. (Sec.~\ref{sec:focusing}, \ref{sec:hadron_dump}, \ref{sec:tagged}), M.P. (Sec.~\ref{sec:proton},~\ref{sec:focusing}), E.G.P. (Sec.~\ref{sec:target}), F.T. (Sec.~\ref{sec:hadron_yields}, \ref{sec:decay_tunnel}, \ref{sec:neutrino_detector}, \ref{sec:conclusions}); validation, writing--review and editing, N.C., A.L., M.P., E.G.P. and F.T. All authors have read and agreed to the published version of the manuscript.}

\funding{This research has received funding from the European Union’s Horizon 2020 Research and Innovation
Programme under Grant Agreement no. 681647 and no. 777419, and by the Italian Ministry for Education and Research
(MIUR, bando FARE, progetto NUTECH). }

\acknowledgments{The authors are grateful to the members of the T2K, DUNE, Hyper-Kamiokande, ENUBET and NA61/SHINE collaborations for many useful discussions and suggestions.}

\conflictsofinterest{The authors declare no conflict of interest.} 




\appendixtitles{no} 


\reftitle{References}


\externalbibliography{yes}
\bibliography{bibliography.bib}




\end{document}